# The Mechanism of the Origin and Development of Lightning from Initiating Event to Initial Breakdown Pulses (v.2)

## Alexander Yu. Kostinskiy[1], Thomas C. Marshall[2], Maribeth Stolzenburg[2]

[1]Moscow Institute of Electronics and Mathematics, National Research University Higher School of Economics, Moscow, Russia, kostinsky@gmail.com

[2]Department of Physics and Astronomy, University of Mississippi, University, Mississippi, USA, marshall@phy.olemiss.edu , mstolzen@phy.olemiss.edu

Corresponding author: Alexander Kostinskiy (kostinsky@gmail.com)

**Key points**:

1. Relativistic runaway electrons from extensive air showers start electron avalanches in many small volumes with electric field > 3 MV/(m·atm)

2. Lightning initiation occurs when many ordinary positive streamers, with speeds of 0.1 - 1 x $10^6$ m/s, develop from these electron avalanches

3. Mechanism accounts for variety in observed characteristics of initiating event, initial electric field change, and initial breakdown pulses

**Abstract**

Based on experimental results of recent years, this article presents a qualitative description of a possible mechanism (termed the Mechanism) covering the main stages of lightning initiation, starting before and including the initiating event, followed by the initial electric field change (IEC), followed by the first few initial breakdown pulses (IBPs). The Mechanism assumes initiation occurs in a region of ~1 $km^3$ with average electric field E>0.3 MV/(m·atm), which contains, because of turbulence, numerous small "$E_{th}$-volumes" of ~$10^{-4}$-$10^{-3}$ $m^3$ with E≥3 MV/(m·atm). The Mechanism allows for lightning initiation by either of two observed types of events: a high power VHF event such as a Narrow Bipolar Event, or a weak VHF event. According to the Mechanism, both types of initiating events are caused by a group of relativistic runaway electron avalanche particles (where the initial electrons are secondary particles of an extensive air shower) passing through many $E_{th}$-volumes, thereby causing the nearly simultaneous launching of many positive streamer flashes. Due to ionization-heating instability, unusual plasma formations (UPFs) appear along the streamers' trajectories of the streamers. These UPFs combine into three-dimensional (3D) networks of hot plasma channels during the IEC, resulting in its observed weak current flow. The subsequent development and combination of two (or more) of these 3D networks of hot plasma channels then causes the first IBP. Each subsequent IBP is caused when another 3D network of hot plasma channels combines with the chain of networks caused by earlier IBPs

## 1. Introduction

Despite great efforts by the scientific community, there is still no generally accepted, qualitatively consistent mechanism of lightning initiation from the initiating event through the subsequent development to the beginning of a stepped leader (e.g., Rakov & Uman, 2003; Gurevich & Zybin, 2001; Dwyer & Uman, 2014). This situation is partly due to the exceptional complexity of the lightning phenomenon, which requires both experimental and theoretical knowledge about lightning itself, along with information from high-energy atmospheric physics, radio physics of atmospheric discharges, physics of turbulent multiphase charged aerosols, gas discharge physics at high pressure, and physics of long sparks. This list could easily be extended. However, after recent significant progress in experimental and theoretical work, there is now an acute need for at least a qualitative construction of a single mechanism describing in space and time the origin and development of lightning.

Based on measurements with a Very High Frequency (VHF) interferometer operating at 20-80 MHz, Rison et al. (2016) "tentatively" concluded that the initiating event of all lightning flashes is a narrow bipolar event (NBE) caused by "fast positive breakdown" (FPB). The NBEs investigated had apparent speeds of 4–10x$10^7$ m/s. Attanasio et al. (2019) recently proposed a FPB propagation mechanism, based on a modernization of the Griffiths and Phelps model (Griffiths & Phelps, 1976) describing initiation of lightning due to a powerful streamer flash from hydrometeors. Using an electric field change sensor (called a "fast antenna" or "FA" with a typical bandwidth of 0.1-2500 kHz), typical isolated NBEs have a characteristic bipolar waveform with a duration of 10-30 μs and large pulse amplitudes (e.g., Willett et al., 1989; Nag et al., 2010; Karunarathne et al., 2015). Typical NBEs also have large power in the HF/VHF frequency band of 3-300 MHz (Le Vine, 1980). For ten positive NBEs initiating intracloud (IC) flashes, Rison et al. (2016) found peak powers in the VHF band (30-300 MHz) ranging from $1 - 274{,}000$ W, while for 5 negative NBEs initiating cloud-to-ground (CG) flashes they found NBE peak powers ranging from $1 - 600$ W. Tilles et al. (2019) further reported that some positive NBEs are caused by "fast negative breakdown" with apparent propagation speeds of 4 x $10^7$ m/s.

Recent findings suggest that most lightning flashes are not initiated by NBEs; rather, most flashes are initiated by much shorter and much weaker events. Marshall et al. (2019) reported the first examples of flashes with much weaker initiating events. Two IC flashes were initiated by VHF events with durations of 1 μs and peak VHF powers of 0.09 W and 0.54 W; there was no coincident pulse in the FA data of these two flashes. Note that the FA data are primarily measuring charge motions with length scale > 50 m while the VHF data are primarily measuring charge motions of length < 5 m, so the lack of an FA pulse with the initiating event suggests charges moved a distance of order 5 m, but not 50 m. Two CG flashes studied by Marshall et al. (2019) were initiated by VHF events with durations of 1 and 2 μs and VHF powers of 0.14 W and 0.64 W; there was a weak, short duration fast antenna pulse coincident with one of the CG VHF initiating events. Lyu et al. (2019) studied 26 IC flashes that occurred within 10 km of their VHF interferometer and found that NBEs initiated only 3 of the 26 flashes; the other 23 flashes were initiated by weak VHF events with durations of less than 0.5 µs. Bandara et al. (2019) investigated 868 negative CG flashes at ranges of 17 – 125 km and found that only 33 (4%) were initiated by negative NBEs; these relatively weak negative NBEs had VHF powers in the range 1-1300 W.

In this article, we describe, as a first step, the main stages of a possible Mechanism for the initiation and development of lightning from the Initiating event through the first several classic initial breakdown pulses. We recognize from the outset the riskiness of such an endeavor, since many processes and phenomena that form the basis of the Mechanism herein proposed by us have not been studied sufficiently or have not yet been considered in such close relationship with each other. This position of limited knowledge gives considerable space for theoretical speculation. But, in our opinion, the construction of a unified Mechanism composed of a consistent sequence of events also has advantages, as it allows future research to focus on quantitative analysis and improvement of (or substantial changes to) each step of the sequence. Thus, our intent is to improve the understanding not only of separate aspects of lightning development, but also of the whole process that combines these key aspects.

For reference, we provide a partial list of terms and abbreviations used in this article:

t1. The ***initiating event*** (**IE**) of a lightning flash is the first electromagnetic manifestation of initiation and can be of the **Weak** sort described by Marshall et al. (2014a, 2019) and Lyu et al. (2019) or of the stronger **NBE** sort as described by Rison et al. (2016) and Lyu et al. (2019). As introduced above, weak IEs have VHF powers < 1 W and durations ≤ 1 µs, while NBEs have orders of magnitude stronger VHF powers and durations of 10 – 30 µs.

t2. A ***Narrow Bipolar Event*** **(NBE)** is a particular type of electrical event that occurs in or near thunderstorms (Le Vine, 1980). (Note that synonyms for "NBE" include "CID" or Compact Intracloud Discharge and "NBP" or Narrow Bipolar Pulse.) An NBE in FA data has a bipolar waveform with a duration of 10-30 µs; in the VHF band of 60 – 66 MHz NBEs have a large power (30,000 – 300,000 W or 45 – 55 dBW) (Rison et al., 2016). The FA data for **weak NBEs** have smaller amplitudes than NBEs and can have either bipolar or "mostly monopolar" waveforms; weak NBEs also have smaller VHF powers of 3 – 300 W or 5 – 25 dBW (Rison et al., 2016).

t3. An ***initial electric-field change*** (**IEC**), as described by Marshall et al. (2014a) and Chapman et al. (2017), is a relatively long period (40-9800 µs) that begins with the IE and ends with the first classic initial breakdown pulse. Marshall et al. (2019) showed that during the IEC there are many VHF pulses with durations of 1 – 7 µs and that some coincident pairs of fast antenna pulses and VHF pulses seem to increase the IEC (as "enhancing events").

t4. An ***initial breakdown pulse*** (**IB pulse or IBP**) is a bipolar electrical pulse occurring in the first few ms of a flash, typically detected with a FA (e.g., Weidman & Krider, 1979; Nag et al., 2009). The largest IBPs are called "**classic IBPs**" and are systematically accompanied by VHF pulses in CG flashes (Kolmašová et al., 2019). By our definition classic IBPs have durations ≥ 10 µs, amplitudes ≥ 25% of the largest IBP, and often have subpulses. Essentially all lightning flashes have a series of IBPs (Marshall et al., 2014b) that occur for a few ms after the IEC; we call the period during which IBPs occur the "**IB stage**" of the flash. During the IB stage, bipolar pulses smaller in amplitude or shorter in duration than classic IBPs are also IBPs; classic and smaller IBPs may be caused by different processes.

t5. A *streamer* is a cold plasma, as described, for example, by Raizer (1991). In this article the term streamer means only "ordinary" streamers, which have been observed for many decades in gas discharges and long sparks at pressures of 1-0.3 atm and have a length from centimeters to several meters (Raizer, 1991, pp. 326-338).

t6. An *unusual plasma formation (UPF)* is a short hot plasma channel as described by Kostinskiy et al. (2015a, 2015b). UPFs often appear as a network of hot plasma channels that are tens of centimeters long.

t7. A *positive leader* is a hot plasma channel that meets certain conditions of length and ambient electric field such that the leader will be self-propagating as described by Bazelyan et al. (2007a). Note that with a sufficient electric field of 0.45 – 0.50 MV/(m atm) (depending on humidity), a hot plasma channel of any length will be self-propagating.

t8. An *EE-volume* (0.1-1 $km^3$) is a region in the thundercloud with average electric field magnitude E>0.28-0.35 MV/(m·atm) and with a large number of charged hydrometeors of different sizes. Hydrometeors can be liquid or solid state, large or small in size, as long as they are plentiful and carry significant electrical charges such that turbulent motions can result in small-scale regions of the EE-volume with substantially larger electric fields. In particular, an $E_{th}$-*volume (or "air electrode")* is a region in the thundercloud with E>3 MV/(m·atm); E magnitudes this large are sufficient to produce "classic" electron avalanches, which, when fulfilling the Meek's criterion (Raizer, 1991), can transform into classical gas-discharge streamers. The EE volume can have strongly inhomogeneous electric fields (on a scale of hundreds of meters) and consist of many closely spaced turbulent regions that can be formed by similarly or oppositely charged countercurrent air flows (e.g., Karunarathna et al., 2015; Yuter & Houze, 1995).

t9. An *EAS-RREA* (extensive air shower — relativistic runaway electron avalanche) (e.g., Gurevich & Zybin, 2001; Dwyer, 2003) occurs when a flow of secondary charged particles of the EAS enters a region hundred meters on a side with electric field E> 280 kV/(m·atm). For the problem of lightning initiation, EASs with primary particle energies $\varepsilon_0 \geq 10^{15}$ eV are important (as described later).

## 2. Experimental and theoretical basis of the Mechanism

The Mechanism proposed herein, despite its complexity, is determined and regulated by reliably established experimental and theoretical work. In this section we list (i1, i2…) the main observations and theoretical ideas that are considered in the development of our Mechanism.

i1. As introduced above, Rison et al. (2016) used an interferometer to detect VHF radiation during lightning initiation, and they located sources at a rate of roughly one per μs. Three positive NBEs that were IEs of three IC flashes had durations of 10-20 μs and very short, exponentially growing fronts of increasing VHF activity with durations of 1-3 μs. The NBE radiation sources advanced downward with apparent speeds of $4\text{-}10\times10^7$ m/s over distances of 500 - 600 m. Bandara et al. (2019) found 33 of 868 CG flashes were initiated by weak negative NBEs, with VHF powers of 1 – 1300 W or 0 – 31 dBW.

i2. Also as introduced above, Marshall et al. (2019) showed that the IEs of two negative CG flashes and two IC flashes were associated with a weak VHF pulse, not an NBE, having a duration of about 1 μs and a VHF power < 1 W. Lyu et al. (2019) showed that 23 IC flashes had an IE that was associated with a weak VHF pulse, not an NBE, with a duration ≤ 0.5 μs.

i3. Marshall et al. (2014b) studied the initiation of 18 CG flashes and 18 IC flashes and showed that for each of the 36 IEs there was no significant electrical activity for 100 - 300 ms before each flash. After the IE an IEC occurred in each flash. Chapman et al. (2017) found IEC durations averaged 230 μs for 17 CG flashes (range 80–540 μs) and 2700 μs for 55 normal IC flashes (range 40–9800 μs), and some flashes had multiple IECs. The physical process causing an IEC is unknown, but apparently the effect is to separate and accumulate enough charge to cause the first classic IBP.

i4. Classic IBPs have range-normalized (to 100 km) amplitudes averaging about 1 V/m (Smith et al., 2018) and estimated peak currents of 1-165 kA (Betz et al., 2008; Karunarathne et al., 2014; N. Karunarathne et al., 2020). High-speed video cameras reveal that there is a bright burst of light with each classic IBP (Stolzenburg et al., 2013, 2014; Campos & Saba, 2013). Stolzenburg et al. (2013) showed and described the light coincident with several series of IBPs in CG flashes as follows: "linear segments visibly advance away from the first light burst for 55–200 μs, then the entire length dims, then the luminosity sequence repeats along the same path" with total lengths of 300-1500 m during the IB stage. These bursts of light indicate the rapid appearance (in less than 20 μs, the frame rate of the camera) of hot, highly conductive channels which mostly vanish in 40-100 μs; after 2-5 ms the IBPs transition to a negative stepped leader with much weaker luminosity (Stolzenburg et al., 2013).

i5. Gurevich et al. (1992) and Gurevich & Zybin (2001) theoretically predicted the important role of cosmic rays in the initiation of lightning: in an electric field E > 218 kV/(m·atm) cosmic rays can cause avalanches of runaway electrons. Gurevich et al. (1999) suggested that the combined action of EAS and runaway electrons could play a significant role in initiating the first streamer in a thundercloud. Dwyer (2003, 2007) introduced another mechanism for generating runaway electrons in electric fields E> 284 kV/(m·atm). For this latter mechanism, positrons and energetic photons produce a positive feedback effect that exponentially increases the number of runaway electron avalanches (Dwyer, 2003, 2007). The Dwyer (2003, 2007) mechanism requires much larger average electric fields than the mechanism of Gurevich et al. (1992, 1999).

i6. Using a balloon-borne electric field meter inside an active thunderstorm, Marshall et al. (2005) estimated that the region where three CG flashes initiated had an average electric field E > 284-350 kV/(m·atm) and occupied a volume of 1 - 4 km$^3$ with vertical and horizontal extents of 300 – 1000 m. This volume is an experimental example of an EE-volume, as defined above. Based on the first detected VHF source of each flash, the three initiations occurred within 1.1 km of the balloon. Marshall et al. (2005) found that the in-cloud E exceeded the relativistic runaway electron avalanche threshold of 284 kV/(m·atm) for about 100 seconds before one of the lightning flashes.

i7. Bazelyan & Raizer (1998, 2000), Bazelyan et al. (2007a), and Popov (2009) theoretically showed the key role of the "ionization-heating instability" in transforming the cold plasma of positive streamers into a hot plasma channel of a long spark leader (i.e., the streamer-leader transition); this transformation (depending on the current strength) occurs in less than ~0.2-0.5 µs, at a pressure of one atmosphere (Popov, 2009; da Silva & Pasko, 2013). If the concentration of neutral molecules in the atmosphere $n$ decreases with height, then, according to recent theoretical calculations, the development time of ionization-overheating instability varies in proportion to $n^{-\alpha}$, where $\alpha$ is in the range from 1 to 2 (Bazelyan et al. 2007b; Riousset et al., 2010; da Silva & Pasko, 2012, 2013). We will use the value $\alpha \approx 1$ (Velikhov et. al, 1977; Raizer, 1991, p. 223-227; Riousset et al., 2010) in our estimates, since there is no experimental confirmation of such high values as $\alpha \approx 2$. Herein we will use the term "ionization-heating instability," although this concept is called different terms in different scientific fields, including "thermal instability" (Raizer, 1991; Nighan, 1977) "ionization-overheating instability" (Panchenko et al., 2006) and "thermal ionizational instability" (da Silva & Pasko, 2012).

i8. Kostinskiy et al. (2015a, 2015b) experimentally showed that in electric fields of 500-1000 kV/(m·atm) within artificially charged aerosol clouds, UPFs are actively initiated, along with bidirectional leaders of 1-3 m length. It was also experimentally shown that in electric fields of 500-1000 kV/(m·atm), UPFs are generated from the plasma of positive streamer flashes via the ionization-heating instability (Kostinskiy et al., 2019).

i9. Colgate (1967) suggested that turbulence in a thundercloud can significantly enhance the local E on scales of about 100 m. Trakhtengerts with co-authors theoretically showed that, due to hydrodynamic instabilities, E in a thunderstorm can fluctuate over about 100 m (Trakhtengerts, 1989; Trakhtengerts et al., 1997; Mareev et al., 1999; Iudin et al., 2003). Trakhtengerts and Iudin (2005) and Iudin (2017) theoretically estimated that more significant amplifications of E on a smaller scale (10-100 cm) are possible due to the statistical movement of hydrometeors of different sizes and charges in a cloud.

## 3. Conditions and Phenomena which the Mechanism Should Satisfy and Explain

Based on the above experimental and theoretical results, the Mechanism must satisfy the following conditions (c1, c2…) and consistently explain the following phenomena:

c1. Overall, the Mechanism should explain how lightning initiation works. The development of the lightning flash should begin immediately after the appearance of the IE. In particular, the Mechanism should explain the series of lightning initiation stages: the IE, the IEC, and the first few classic IBPs in the IB stage. These three initial stages of a flash are followed by the well-understood negative stepped leader stage.

c2. The optical radiation of IEs seems to be quite weak (Stolzenburg et al., 2014, 2020). Thus the Mechanism should not contain an initial powerful flash of light like the powerful burst of light that occurs during an IBP.

c3. For both types of IEs (NBE or Weak): during the IEC the Mechanism should develop conducting paths of several kilometers so that significant charge can be stored in corona sheaths and so that charge can flow, thereby producing the IEC.

c4. For IEs that are NBEs (e.g., Rison et al., 2016): the Mechanism should explain the production of the NBE itself, including a short, powerful flash of positive streamers with an exponentially increasing radiation risetime of a few microseconds, a total duration of 10–30 μs, and a very strong VHF signal.

c5. For IEs that are NBEs: Before and during the powerful VHF radiation that NBEs produce, the Mechanism must not produce a strongly emitting long hot conductive plasma channel (Rison et al., 2016).

c6. For IEs that are NBEs: the Mechanism should contain a physical process that moves at a speed close to the speed of light to match the experimental data (Rison et al., 2016) while also providing a short duration narrow bipolar pulse in FA data.

c7. The Mechanism should not contradict the well-known and well-tested data on gas discharge physics and the physics of a long spark: e.g., propagation speed of positive streamers depending on the electric field; avalanche-streamer transition; streamer-leader transition; fast attachment of electrons to oxygen molecules, etc. In particular, laboratory measurements show that the speed of positive streamers is usually $1 - 10$ x $10^5$ m/s with a maximum of 5 x $10^6$ m/s for E of 3-4 MV/(m·atm) (Les Renardieres Group, 1977).

c8. The Mechanism should explain how the first classic IBP (with a current ≥ 10 kA and a bright light burst) is produced so soon after the IE.

## 4. Some Main Components of the Mechanism

### 4.1. The IE, EAS-RREA, and avalanche-streamer transition

The Mechanism assumes that the IE consists of a three-dimensional (3D) group of classic electron avalanches rather than a single avalanche. A "classic" electron avalanche develops in an electric field E ≥ 3 MV/(m·atm). The Mechanism further assumes that most of the classic electron avalanches in the group are started by an electron (or several electrons) freed from atoms by a relativistic electron, relativistic positron, or high energy photon of a relativistic runaway avalanche (Gurevich & Zybin, 2001; Dwyer, 2003), which occurs when a EAS ($\varepsilon_0 > 10^{15}$ eV) occurs in the region of a strong electric field E > 0.4 MV/(m·atm)

In order for an avalanche to transform into a positive streamer (i.e., undergo avalanche-streamer transition), the avalanche must produce about $10^8$-$10^9$ electrons in a volume of about 0.3-0.5 mm$^{-3}$. This is Meek's criterion (Raizer, 1991). When this electron density is reached (2-3·$10^8$-$10^9$ mm$^{-3}$), a strong E, created by the polarization of the avalanche head by the cloud E, starts a self-sustaining ionization in front of the head. Accordingly, the E of the streamer head begins to exceed several times 3 MV/(m·atm). After that, the streamer begins to move independently. This process is indicated by a rough estimate of the electric field of a sphere of such dimensions. $E = \frac{1}{4\pi\varepsilon_0} \frac{Q}{R^2}$, $E \approx 9 \cdot 10^9 \cdot (1.6 \cdot 10^{-19}) \cdot 10^9 / (0.4 - 0.2 \cdot 10^{-3})^2 \approx 9$-36 MV/m. But in order for the streamer to be fully formed from one or several initial electrons, E > 3 (MV/(m·atm)) is necessary over the entire length of the avalanche growth. This length is determined by Meek's criterion $\alpha_{eff} \cdot d \approx 20$, where $\alpha_{eff}$ is the Townsend coefficient, which is 10-12 cm$^{-1}$ for air at 1-atm pressure and with $E$ of ≈ 3.0-3.2 MV/(m·atm) (Raizer, 1991). This means that the length of the avalanche when it transforms into a streamer will be about 2 cm at atmospheric pressure (Raizer, 1991) and this length will increase exponentially with height. See Supplemental Material S1 for more discussion of the avalanche-positive streamer transition.

## 4.2. Positive streamer flashes

Positive streamer flashes are comprised of a large number of positive streamers; positive streamer flashes that start from metal electrodes have been well studied. The task of determining the parameters of a streamer flash in a thundercloud at the time of lightning initiation limits the relevant experimental data to the following parameters: voltage and E risetimes (or fronts) must be > 100 μs, and the size of the "air electrodes" in the cloud must be ≥2-5 cm. We give a detailed explanation of these restrictions in the next section.

### 4.2.1. The front and duration of a typical individual positive streamer flash

Figure 1(a) shows a typical current oscillogram of a streamer flash on the electrodes with a diameter of ~5-25 cm in discharge gaps of 4-20 m (Les Renardieres Group, 1977, Bazelyan & Raizer, 1998). Figure 1(b) shows the corresponding image-converter picture of the positive streamer flashes emitted from the electrode. The voltage pulse causing the positive streamer flashes had a rather slow front of ~200-500 μs, and a pulse duration of 2.5-10 ms.

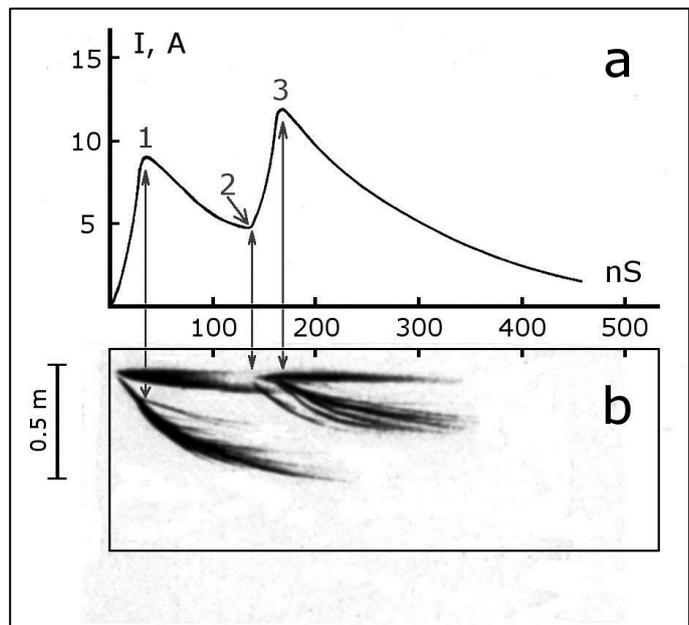

Figure 1 (sketch). A typical streamer flash starting from a metal electrode. The diameter of the electrode is ~5-25 cm. The voltage front is ~200-500 μs, the rod-plane gap is 4-20 m. (a) current at the HV electrode; (b) simultaneous image-converter picture.

In such streamer flashes from the electrode (Figure 1), the leading front of current growth is ~25–40 ns, and it grows until the first branching of streamers (1). At time (2), the second branching of streamers begins (for the second streamer flash) with a front of ~30-60 ns (3). The whole streamer flash lasts about 300-500 ns. The voltage at which the streamer flash is initiated depends strongly on the electrode dimensions and its capacity. For example, for a rod electrode with a hemisphere cap 100 mm in diameter, the initiation voltage of the streamers is ≅509 kV (and E on the electrode surface reaches ≅6 MV/m), but for a sphere with 1 m diameter, the voltage reaches ≅1855 kV (and E on the electrode surface is ≅3.2 MV/m), Les Renardieres Group, 1977. Also, the size of the electrode significantly affects the total charge of the streamer flash: for a hemisphere with a diameter of 100 mm, total charge is 6.8 ± 3.4 μC, while for a sphere of 1 m diameter, total charge is 62 ± 1 μC. In clouds, due to the relatively long lifetime of the electric fields, the generation of streamers will most likely take place in E close to the breakdown field of 3 MV/(m·atm), without a large overvoltage and with a very slow voltage rise compared to discharges on electrodes. Kostinskiy et al. (2015a, 2015b) studied positive streamer flashes in a charged aerosol cloud, and these experiments are useful for our understanding of positive streamer flashes in a thundercloud. When generating a charged aerosol cloud, the rise of the voltage front on a grounded sphere with a diameter of 5 cm was 300-500 ms, and the duration of the applied voltage was several tens of minutes, which is closer to the actual conditions in a thundercloud.

Therefore, we describe the positive streamer flashes in the E of an aerosol cloud in detail and show their development in Figure 2 (Kostinskiy et al., 2019), since understanding their development is useful in constructing the Mechanism. It is important that even in the conditions of a very slowly varying E (e.g., over 300-500 ms) created by a charged aerosol cloud, the current fronts (risetimes) of streamer flashes also had a duration of about 30 ns as in the experiment described above and shown in Figure 1. The first streamer flash of Figure 2a(1) had a current peak of 1.1 A, the current front was 30 ± 5 ns, the half-width of the peak at half-height was 90 ± 10 ns, and the falltime was 147 ± 10 ns. A picture of this streamer flash is shown in Figure 2b.I(1). The total duration of the flash current was about 200 ns. Streamers continued to exist and move even after the current on the electrode dropped to zero, since E in the entire region from sphere (2) to cloud (3) exceeded the threshold required for movement of positive streamers in air of $E_{str+} \geq 0.45$-0.5 MV/(m·atm) (Bazelyan & Raizer, 1998). Streamers flew to the center of the cloud in 1.7 μs. Therefore, the duration of the streamer flash current measured at the electrode (≈ 200 ns) is determined by the time of loss of the galvanic (current) connection of the streamers with the metal sphere. The streamers themselves moved and existed for at least 1.7 μs. The flash had a length of at least 1.2 meters and the streamers moved from the sphere, Figure 2b.I(2), to the area labeled (5), Figure 2b.I, where they were detected by a microwave diagnostic beam. Since the streamers were moving and did not transition to positive leaders, we know that the E values in their path were > 0.5 MV/m and < 1 MV/m. The average speed of the positive streamers of the first flash in the image plane was found to be ≈ 7x10$^5$ m/s. The small maxima and minima of the current on the oscillogram of the first streamer flash Figure 2a(6), as well as on the entire oscillogram of the current, are artifacts of the measuring circuit and should be ignored. The two other current maxima, which correspond to the second and third streamer flashes of Figure 2a(2,3), had similar current risetimes (≈ 30 ns) as the first streamer flash, despite having much higher currents (3.14 A, 5.8 A). Thus, the risetime of a streamer flash in such E values does not change and characterizes the physical parameters of individual streamers. The total current of an individual streamer flash from a large-diameter electrode can reach 100–200 A, but the duration of the leading front of the streamer flash varies within small limits of 25–35 ns. Such a rapid front yields powerful VHF radiation with a maximum in the 30-40 MHz range. The maximum current in Figure 2a(4) is most likely due to ionization-heating instability, which caused the leader to appear, Figure 2b(7). The front of the leader has different parameters: risetime of 195 ± 10 ns (6 times slower than the first three positive streamer flashes (1,2,3)), half-width at half maximum of 180±10 ns, and falltime of 210±10 ns. This positive leader will have a peak emission frequency of 5 MHz and a duration 0.4 μs and so would probably be detectable with the FA (described above). Later in the development, the small peaks with similar parameters correspond to the current of the positive upward leaders (Figure 2a(5)), which move by small steps (jumps) that are a few cm long.

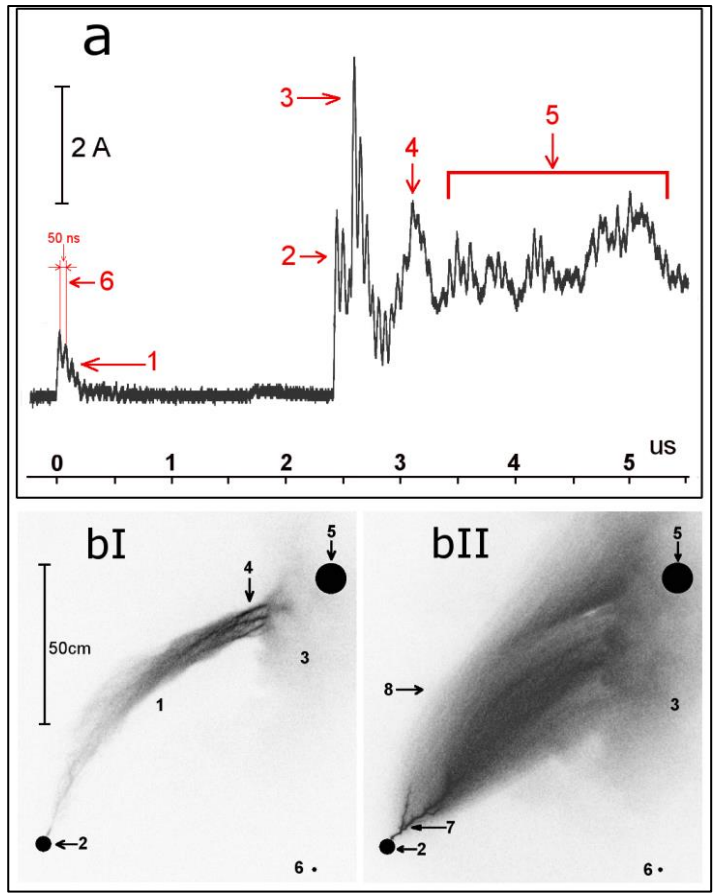

Figure 2. (**a**). A waveform of a typical positive streamer flash from an electrode without overvoltage. The rise of the voltage front on a grounded ball with a diameter of 5 cm was 300-500 ms, and the duration of the applied voltage is a minute. Electric field on the electrode ≈3 MV/m. Numbered features: 1, 2, 3 — first, second and third streamer flashes, 4 — leader initiation current, 5 — moving leader current. (**b**). Two consecutive images of the streamer flash whose data are shown in Figure 2a. The streamer flash started from a grounded electrode (2) in the electric field of a charged aerosol cloud. The two images were taken with a 4Picos camera with image enhancement: bI —first frame with 2 μs exposure; bII — second frame with 10 μs exposure; time interval between frames 1 μs, both frames are inverted. Numbered features: 1 - the first flash of positive streamers; 2 — 5 cm grounded sphere equipped with current-measuring shunt; 3 — cloud of charged water droplets; 4 — UPFs; 5 — the area of passage of the microwave beam; 6 — the center of the grounded plane where the nozzle is located; 7 — upward positive leader; 8 — streamer crown of a positive leader *(Adapted from Kostinskiy et al, 2019).*

### 4.2.2. Length and conductivity of long streamers

For long streamers the movement of the streamer head is supported by its own strong E, which ionizes the air in front of the head. This E considerably exceeds $E_{th}$ required for Different estimates of the E in front of the streamer head are in the range 10–30 MV/(m·atm), Bazelyan and Raizer (1998, 2000). When the head crosses this point of its trajectory (i.e., the location with E of 10-30 MV/(m atm)), then E immediately drops to a value close to the ambient E. In this case, the ionization frequency very rapidly decreases, while the frequency of electron attachment to oxygen molecules begins to play the main role, (Bazelyan & Raizer, 1998, p.25). At E < 0.5 MV/(m·atm), there is a strong loss of

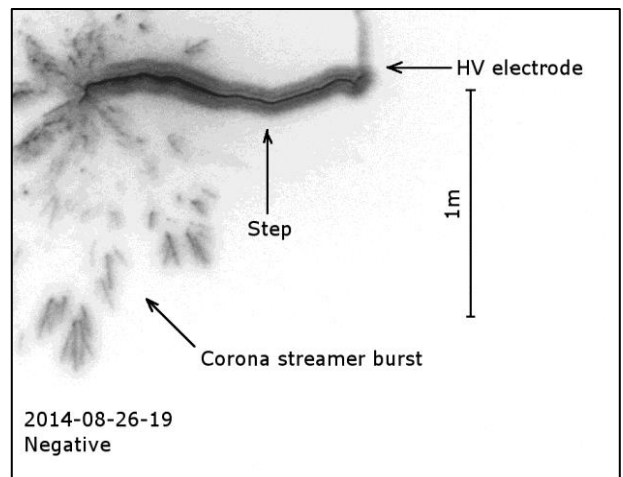

Figure 3. Event 2014-08-26-19. Negative leader step with pronounced corona streamer burst. Only the first of the two 4Picos frames is shown (the second one may contain a camera artifact). Exposure time was 50 ns. Note that streamer branches appear to extend in all possible directions, although this could be in part due to unusual channel geometry. This leader did not reach the opposite electrode. The focal length was 50 mm, and the f-stop (relative aperture) was f/0.95. The size of image pixel in the object plane is 3.1 × 3.1 mm$^2$. The ambient temperature was 11°C, and the humidity was 81% (Adapted from *Kostinsky et al. 2018*).

plasma conductivity behind the streamer head in only 100–200 ns at a pressure of 0.3–1 atm. In typical E values of a thundercloud, the main loss mechanism is the three-particle attachment of electrons to oxygen molecules (Kossyi et al., 1992, Bazelyan & Raizer, 2000):

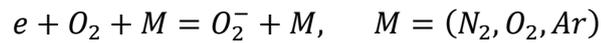

$$e + O_2 + M = O_2^- + M, \quad M = (N_2, O_2, Ar)$$

with the attachment rate falling in proportion to the square of pressure. Thus, if the propagation velocity of streamers is 2-5x10$^7$ cm/s, and the characteristic time of attachment is 1-2x10$^{-7}$ s, then the length of the streamer conductive channel at atmospheric pressure will be 2-10 cm. The concentration of electrons in the streamer head reaches 1-3x10$^{14}$ cm$^{-3}$, and behind the streamer head the plasma concentration decreases exponentially with the length (Bazelyan & Raizer, 1998, 2000). The near-real length of streamers at atmospheric pressure is visible in Figure 3 (Kostinsky et al., 2018) and is 10–20 cm. The physical nature of electron attachment to oxygen does not change for positive and negative streamers. At an altitude of 6 km above the Earth, where the pressure has decreased by a factor of 2, the length of the conductive channel of the streamers will increase four times and reach a length of 8-40 cm (depending on E), and at an altitude of 9-10 km the length of the streamers will be 18-90 cm with the same exponential decrease in conductivity behind the streamer head.

### 4.3. The IE as a nearly simultaneous initiation of a group of positive streamer flashes in a thundercloud

In the previous section, we described individual streamer flashes that occur on a metal electrode, but (in nearly all cases) there are no metal electrodes in a thunderstorm cloud. In a thundercloud there are various hydrometeors that are statistically distributed within the cloud and carry different charges. We assume that the lightning Initiating Event produces a system of positive streamers distributed in space and time, with the minimum system being a single positive streamer.

### 4.3.1. Corona from hydrometeors producing the group of positive streamer flashes

One possibility for lightning initiation may be that positive streamers are emitted from one or more hydrometeors (e.g., Dawson & Duff, 1970; Phelps, 1974; Griffiths & Phelps, 1976). For example, a system of hydrometeors distributed in a volume can generate a system of positive streamer flashes distributed in the volume. Thus, if the IE is an NBE, a large total streamer flash may, as suggested by Rison et al. (2016), consist of a set of individual flashes initiated by hydrometeors and distributed in a volume; each of these flashes may be similar to the streamer flashes from metal electrodes described above. If the IE is a weak event (e.g., Marshall et al., 2019), it may also be a 3D set of positive streamer flashes but with fewer individual streamers. Electromagnetic radiation from the volumetrically and temporally distributed positive streamers, as detected by instruments at the ground, may appear as point sources developing in 3D space and in time. Thus, the combined event consisting of a set of individual streamer flashes should generate the commonly observed, single VHF pulse of the IE.

For NBEs Rison et al. (2016) stated "*the breakdown appears to be produced by a spatially and temporally distributed system of positive streamers, in which the total current is spread over some cross-sectional area as a volume current density*." Rison et al. (2016) estimated a spatial scale of ~500 m for NBEs and suggested that the positive "*streamers would be initiated by corona from ice crystals or liquid hydrometeors.*" However, each streamer initiation requires a thermal free electron (energy < 100 eV) near the initiating-hydrometeor (Dubinova et al., 2015; Rutjes et al., 2019), and modeling suggests that the thermal free electrons may not be available over spatial scales of 500 m x 500 m (Dubinova et al., 2015; Rutjes et al., 2019). Some studies have inferred that the initiating-hydrometeor size must be larger than believable, 6 – 20 cm (e.g., Dubinova et al., 2015; Babich et al., 2016). These problems with the hydrometeor-corona IE process encouraged us to consider an alternative IE process, as described in the next paragraph.

### 4.3.2. Hydrodynamic and statistical processes for enhancing the electric fields of a thundercloud and thereby producing the group of positive streamer flashes

Another possibility for the IE of a lightning flash (as well as for an isolated NBE) is that hydrodynamic and statistical processes (driven by turbulence) could enhance E in many small regions of the cloud to allow individual free electrons to produce a 3D set of positive streamer flashes; this set would then radiate in the VHF-band, as discussed in the previous section. Thus, this alternative way of producing a group of positive streamer flashes would account for the measurements of lightning initiation and development. The small-scale E enhancements probably require a relatively large region of relatively strong E, which seems to exist in thunderclouds, as discussed next.

Based on balloon soundings through active thunderclouds, the typical large-scale charge distribution has 4 – 8 horizontally extensive charge layers distributed vertically with typical maximum measured vertical E of ±350 kV/(m atm) (Marshall & Rust, 1991; Stolzenburg & Marshall, 2009). Local maximum E magnitudes typically occur between opposite polarity charge regions, so these maxima are distributed vertically through the cloud. Less is known about the volume of the large E regions, but Marshall et al. (2005) used balloon E measurements and found volumes of 1 - 4 km$^3$ with vertical and horizontal extents of at least 300 – 1000 m associated with three lightning flash initiations. Note that the balloon studies mentioned above ignored small-scale E variations for cloud depths < 100 m and short duration E variations for times < 10 s. However, Stolzenburg et al. (2007) studied 9 balloon flights in which the balloon and/or instruments were struck by lightning and found maximum E magnitudes just before seven lightning strikes of 309 – 626 kV/(m atm) and inferred magnitudes of 833 and 929 kV/(m atm) before the other two strikes. Another finding in Stolzenburg et al. (2007) supports the turbulence-driven, small-scale enhancement of thundercloud E underlying our Mechanism, especially since the typical duration of a CG lightning flash is 200-300 ms (Table 1.1, Rakov & Uman, 2003): Stolzenburg et al. (2007) reported that for 7 of the 9 lightning strikes, |E| increased rapidly for 2-5 s before the lightning strikes with dE/dt magnitudes of 11-100 kV/m/s. A few seconds before one of the flashes there was a step increase (in < 1 s) in measured E of 380 kV/(m atm) that lasted for only 1 s and before another flash there was a step increase of 505 kV/(m atm) that lasted 13 s, then E declined by 15% for the last 2 s before the flash occurred. Thus the E data in Stolzenburg and Marshall (2009), Marshall et al. (2005), and Stolzenburg et al. (2007) support the existence of EE-volumes within thunderclouds. These works are also consistent with the idea of smaller-scale regions occurring for short times with much larger E values caused by the hydrodynamic, turbulent, and statistical nature of the charge distribution in a thundercloud.

It is our hypothesis that discharges in a thundercloud are fundamentally different from the laboratory discharges from metal electrodes precisely because the E is created by statistically located charges in the cloud, which create small-scale variations in E. These small-scale variations of the field do not exist in the physics of a classical electrode gas discharge. For several decades, an approach has been developed for the possible local enhancement of E in a thunderstorm on scales from tens of centimeters to hundreds of meters. As introduced above (i9), fluctuations of E over about 100 m can occur due to hydrodynamic instabilities (Colgate, 1967, Trakhtengerts, 1989; Trakhtengerts et al., 1997; Mareev et al., 1999; Iudin et al., 2003). Furthermore, Trakhtengerts and Iudin (2005) and Iudin (2017) have shown that additional E enhancements at even smaller scales may be possible due to the random statistical motion of a multitude of charged particles of different sizes in the cloud. Unfortunately, these approaches rely on theoretical work, and it is difficult to verify them experimentally. The calculations so far have not incorporated the full structure of real turbulence in the cloud and the full dynamics of real hydrometeors due to the very large computational complexities. Brothers et al. (2018) examine this topic using a large-eddy-resolving model (125 m grid). Their numerical simulations for two storms showed "tremendous amounts of texture," or small-scale spatial variations and inhomogeneities, in the charge density due to charge advection in large-eddy turbulence. Although they do not show the small-scale E, Brothers et al. (2018) mention that there should be "more favorable locations for breakdown to occur" due to there being more neighboring small pockets of opposite charges. It seems a reasonable conjecture that higher spatial resolution modeling of this sort, to include smaller turbulent eddies along with meter-scale hydrodynamic instability effects, will yield even smaller scale E enhancements.

This hypothesis of small-scale variations in E seems promising to us for two main reasons. First, we have not found any contradictions between this hypothesis and measurements of lightning or E in real thunderstorms. Second, the development of a lightning flash from the IE through the IB pulses is quite varied from flash to flash with a wide range of IEC durations and amplitudes (e.g., Marshall et al., 2014b), a wide range of IB pulse durations, inter-IB pulse times, IB pulse amplitudes, and a wide range in the number of subpulses (0-5) on the classic IBPs (e.g., Marshall et al., 2013; Stolzenburg et al., 2014; Smith et al., 2018). Of special note in the flash development is the seemingly random order of the classic IBPs: in roughly 1/3 of flashes the first classic IBP has the largest amplitude followed by classic IBPs with varying amplitudes, while in many flashes the largest classic IBP is third, fourth, or fifth with a median time after the first IB pulse of 1.4 ms for IC flashes and 0.25 ms for CG flashes (Smith et al., 2018). These variations in flash development from IE through the IB stage are easier to understand in the context of a statistical distribution of small-scale regions of large magnitude E, rather than with a single, smooth region of large E.

### 4.3.3. Advantages of the process of hydrodynamic and statistical enhancement of E.

First, it is important to note that this process for increasing E does not depend on the phase state of the hydrometeors, and therefore it fits with the fact that lightning initiation occurs over the altitude range of 3 – 15 km (or greater). In this process hydrometeors of any size (i.e., cloud particles or precipitation particles) can be liquid and highly conductive or completely frozen with poor conductivity. The main requirement is that hydrometeors carry charge and move in turbulent flows, such that hydrodynamic instabilities in a turbulent cloud will lead to charge density variations and thus to increases and decreases in E on scales of tens of meters or smaller. That is, within an EE-volume of 1000x1000x1000 m or 500x500x500 m and large-scale E ≥ 284-350 kV/(m·atm), there are fluctuations of E with scales of tens of centimeters to tens of meters, and E amplitude variations on the order of 10-100 kV/(m·atm) or more, as we discuss in more detail next.

In the EE-volumes, charged hydrometeors are separated by distances of several millimeters to tens of centimeters and moving randomly. In Brownian motion, a multitude of 0.03 μm molecules can simultaneously strike a large particle 3 μm in size on one side, while a smaller number of molecules hit it on the other side, thereby causing the large particle to make a big jump. One molecule cannot affect a larger particle that is a million times heavier, but in an ensemble there are always fluctuations proportional to $\sim\sqrt{n}$ molecules to produce Brownian motion in a larger particle. Similar processes should also occur in the cloud during random motion among differently charged hydrometeors (Iudin, 2017). This movement of hydrometeors leads to a broad spectrum of amplitude oscillations in E. Rare, small-volume, large-amplitude oscillations of E should be added to larger scale oscillations of hydrodynamic changes in E. These statistical waves of oscillations of E are estimated to have a scale of 1-30 cm (Trakhtengerts & Iudin, 2005; Iudin, 2017). Ordinary metal electrodes have a similar scale (1-30 cm) during high-voltage discharges. Hypothetically, these increases and decreases in E can add up with each other and lead, on a small scale, to a spectrum of E values up to the breakdown value of 3 MV/(m·atm). The volumes with scales of 1-30 cm and E ≈ 3 MV/(m·atm) would be the "air electrodes" mentioned earlier. Of course, such large fluctuations in the thunderstorm should occur very rarely (no more than one small region with E ≈ 3 MV/(m·atm) per 3-100 cubic meters), but unlike high-voltage electrodes, the thundercloud has dimensions in cubic kilometers, and the lifetime of a strong average E may be tens of minutes. It is important that the probability of waiting for such strong amplitude oscillations of E, as in other similar statistical processes, is proportional to $\sim\sqrt{t}$. The cloud is able to "wait patiently" and probably "manages to witness" such strong local oscillations of E in conditions of strong turbulence. The characteristic lifetime of a small-scale field gain should be from tens to hundreds of milliseconds (Iudin, 2017). Moreover, in the typical large field of an EE-volume there should be a multitude of small E increases, creating a network of "mountain ranges" of E with peaks and valleys (depicted in Figure 4). In our Mechanism it is the network of "tops" of the electric field mountain range, exceeding in some places $E_{th}$ of 3 MV/(m·atm) and separated by centimeters to meters to tens of meters, that can initiate positive streamers. In our Mechanism these large E locations ("tops" or $E_{th}$-volumes or "air electrodes") replace the large hydrometeors proposed by others to enhance E at the tips of hydrometeors and thereby start positive streamers (e.g., Dawson & Duff, 1970; Phelps, 1974; Griffiths & Phelps, 1976; Rison et al., 2016). In other words, our Mechanism focuses on the E "landscape" rather than on the hydrometeor distribution.

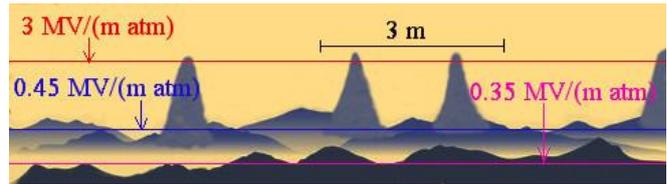

Figure 4. Possible variations of the electric field in turbulent streams (jets) of a thundercloud.

The second advantage of this hydrodynamic and statistical approach is that the dimensions of volumes where E exceeds $E_{th} > 3$ MV/(m·atm) are in the range 2-5 of centimeters (see Kostinskiy et al., 2020), and these dimensions at lightning initiation altitudes of 3 to 15 km will be sufficient for the transition from avalanche to streamer.

If the initial streamer flashes are completely formed in a small number of areas with very strong E, then positive streamers will be able to advance providing the average E exceeds the minimum required of $E_{str+} \geq 0.45$-$0.5$ MV/(m·atm). Transition of streamer flashes into a hot conductive channel will occur when the streamers reach an $E \approx 0.5$-$1.0$ MV/(m·atm). After traveling a distance of several meters, these streamers will, with a high probability, transition into short (1-30 cm), hot, conductive plasma formations similar to UPFs (Kostinskiy et al., 2015 a,b). Statistically, the areas with $E > 500$-$1000$ kV/(m·atm) should be larger than areas with a very high $E > 3$ MV/(m·atm).

## 5. The Mechanism

In this section we describe our Mechanism for lightning initiation from the IE to the negative stepped leader. Our Mechanism is only slightly different for the two cases of (1) the IE is a NBE (either strong or weak) as in Rison et al. (2016), and (2) the IE is a much weaker event as in Marshall et al. (2019). We describe the NBE case first.

### 5.1. The Mechanism for lightning initiation by NBEs, or the NBE-IE Mechanism

An important condition for this case of the Mechanism is condition c6: A short and powerful phenomenon that initiates NBEs should move at a speed close to the speed of light (Rison et al., 2016). For such rapid propagation, ordinary streamers are not suitable. Even in relatively large E, positive streamers will travel only 500 m in 500 μs ($v_{str} \approx 10^6$ m/s) because for motion over such large distances, only the average E is important, so statistical field enhancements and attenuations are averaged. To propagate with a speed close to the speed of light, an electromagnetic pulse seems to require a hot, well-conducting plasma channel, as found for a return stroke or after the end-to-end meeting of two bidirectional leaders (Rakov & Uman, 2003; Jerauld et al., 2007). Even dart leaders moving in previous return stroke channels only attain speeds of one tenth of the speed of light. Thus for NBEs to actually move at a speed close to the speed of light, it seems that hot channels are needed; this implication directly contradicts the experimental data of conditions c5 (before and during the NBE, a long hot conductive plasma channel must not be produced) and c2 (during the IE there should not be a powerful flash of light). Of all the physical processes known to us, the only process moving at a speed close to the speed of light without the assistance of a highly conductive hot plasma channel is the movement of relativistic particles (Gurevich & Zybin, 2001; Dwyer, 2003), so we employ relativistic particles in our Mechanism. (See also Kostinskiy et al., 2020.)

The steps in the NBE-IE Mechanism for flash initiation by an NBE are detailed in the following subsections.

### 5.1.1. Necessary conditions.

The NBE IE will occur in an EE-volume of the thundercloud as defined earlier: a volume of about of 0.1-1 km$^3$ with a high *average* E>0.28-0.35 MV/(m·atm) and a large number of charged hydrometeors of different sizes. Hydrometeors can be liquid or solid, large or small, and a substantial number of them must carry a significant electrical charge. As described earlier (idea i6), Marshall et al. (2005) measured E of this magnitude in volumes this large and larger.

### 5.1.2. Large electric fields develop because of turbulent motions.

In the EE-volume, clouds develop hydrodynamic instabilities and turbulence, which lead to strong local statistical fluctuations of E of different scales and sizes on the background of the average large E: see Figure 5.I. Due to the spectrum of oscillations of E, we assume that the $E_{th}$-volumes ("air electrodes") are relatively small (3-60 cm$^3$) and exist for at least a few tens of milliseconds (Kostinskiy et al., 2020); recall that in $E_{th}$-volumes the local E exceeds the conventional breakdown threshold $E_{th}$ ≥ 3 MV/(m·atm) necessary to start streamers. The small $E_{th}$-volumes are marked with "1" in Figure 5.I and with "6" in Figures 5.II-5.IV. A simple estimate given in the next paragraph shows that in order to provide NBEs with sufficient streamers to explain the total charge moved in an NBE, one $E_{th}$-volume should, on average, be found in each volume of approximately 5m x 5m x 5m. Larger volumes are also formed with an average electric field strength greater than the threshold for propagation of positive streamers $E_{str+}$ ≥ 0.45-0.5 MV/(m·atm), see Figure 5.I(3); we call these regions $E_{str+}$-volumes. Of course, many regions in the EE-volumes will have smaller E than the average E (Figure 5.I(2)).

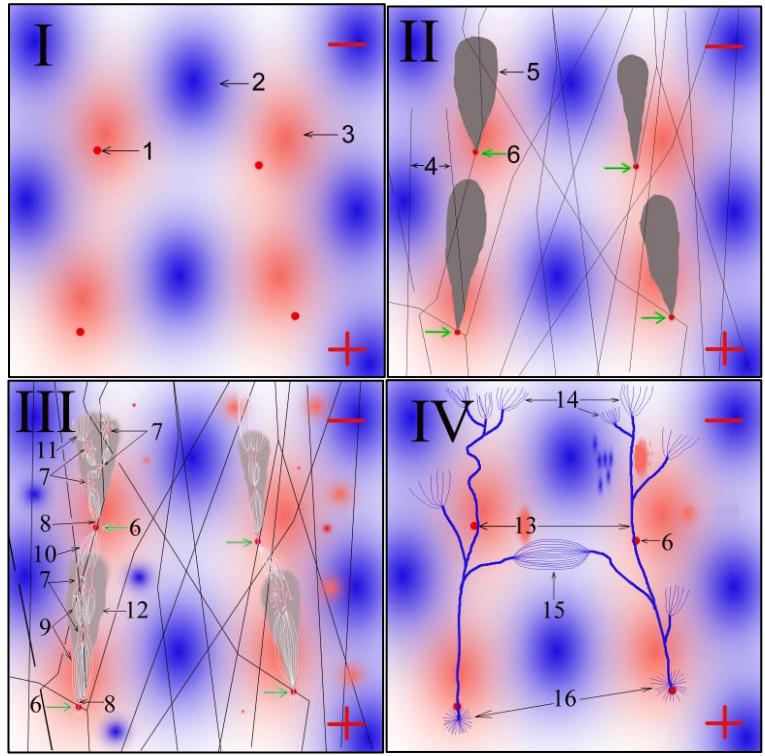

Figure 5. Sketch of the mechanism of initiation of NBEs, which lead to the initiation of lightning. Numbered features: 1,6 - area with an electric field E≥ 3 MV/(m · atm); 2 - area with an electric field E< 0.45-0.5 MV/(m·atm); 3 - $E_{str+}$-volume, area with an electric field E ≥ 0.45-0.5 MV/(m·atm); 4 - runaway electron trajectories; 5 - positive streamer flashes; 7,8 - UPFs; 9 - secondary streamer positive crowns connecting UPFs; 10 - secondary streamer crowns connecting UPFs of two different streamer flashes; 11 - positive crown before UPFs; 12 - the trace of the first streamer flash; 13 - hot highly conductive plasma channels; 14 - streamer crowns of positive leaders; 15 - streamer corona of two interacting large plasma channels; 16 – streamer flash negative leader crown.

We now give a simple estimate of the number of $E_{th}$-volumes needed for a strong NBE. We assume that each $E_{th}$-volume may act as an "air electrode." Let the charge carried by the NBE be about 1 C, since estimated charges of the three NBEs in Rison et al. (2016) were 0.5, 0.7, and 1.0 C. The charge of a small streamer flash from an "air electrode" of about (10 cm)$^3$ (= 10$^{-3}$-10$^{-4}$ m$^3$ = 0.1-1 liter), where $E_{th}$ ≥ 3 MV/(m·atm), Figure 5.III(6) - Figure 5.IV(6), should be similar to a positive streamer discharge in Figure 2.I(1) with a charge of about 10$^{-7}$ C. Thus about 10$^7$ of $E_{th}$ volumes are required to provide a 1-C charge. We can divide an EE-volume of 1000x1000x1000 m$^3$ into 10$^7$ equal volumes that are 100 m$^3$ ($\approx$ 4.6 x 4.6 x 4.6 m). It seems reasonable to imagine that each one of the 100 m$^3$ volumes might contain one region of 10$^{-3}$-10$^{-4}$ m$^3$ with a large statistical enhancement to E $\geq$ 3 MV/(m·atm), since there are very many particles in 100 m$^3$ (e.g., in a thunderstorm anvil, Dye et al. (2007) found concentrations of 10$^4$-5x10$^5$ per m$^3$ for particles with diameters of 30 μm $-$ 2 mm) and measurements suggest that in a thunderstorm many of the cloud particles and precipitation particles are charged (e.g. Marshall & Stolzenburg, 1998).

### 5.1.3. EAS-RREA-synchronized (nearly simultaneous) start of a large number of electron avalanches and streamer flashes

When the necessary conditions in section 5.1.2 are fulfilled, then in order to start streamer flashes in the E$_{th}$-volumes, the EE-volume must be "sown" with ionizing particles (electrons, positrons, photons). The number of nearly simultaneous ionizing particles in the EE-volume must be so large that in a few microseconds the ionizing particles get into most of the E$_{th}$-volumes: see Figure 5.II(6). The ionizing particles will produce thermal free electrons needed to start classic electron avalanches. For this, EASs ($\varepsilon_0 >$ 10$^{15}$ eV) are ideal; in strong E> 0.284 MV/(m·atm) they will generate avalanches of runaway electrons (Gurevich & Zybin, 2001; Gurevich et al. 1999; Dwyer, 2003). The general scheme of the process of synchronization of streamer flashes due to the EAS-RREA mechanism, for example, between an upper negative and a lower positive charge is shown in Figure 6. A cosmic ray particle with an energy $\varepsilon_0 >$ 10$^{15}$ eV (labeled 1 in Figure 6) creates EAS (2). EAS electrons and positrons (3) enter the region of a strong electric field (5), which can support the propagation of RREA (4) and positive streamers (7). EAS-RREA electrons and positrons (labeled 6 in Figure 6) cross the region of a thundercloud with air electrodes (9) and synchronize the nearly simultaneous triggering of multiple streamer flashes (7), starting within the air electrodes through which the electrons pass (labeled 8 in Figure 6). Numerical simulation of this process is given in Kostinskiy et al. (2020).

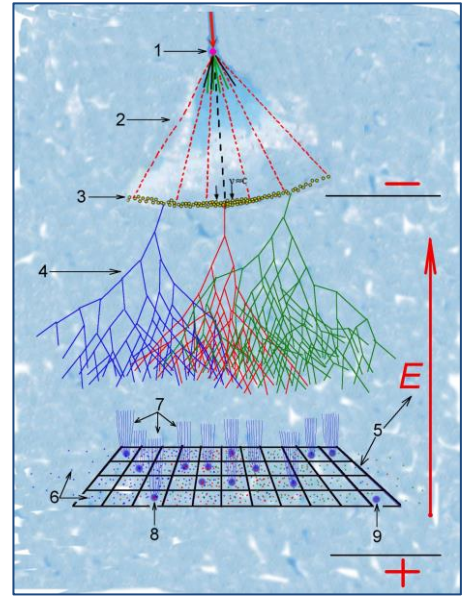

Figure 6. Schematic depiction of Energetic Air Shower (EAS). 1 - the primary particle of EAS; 2 - EAS; 3 - secondary EAS electrons; 4 - RREA; 5 - region of a strong electric field; 6 - EAS-RREA electrons crossing the region of strong turbulence of a thundercloud, which creates strong electric fields; 7 - EAS-RREA synchronized streamer flashes; 8 - an air electrode (E$_{th}$-volume) through which an energetic electron has passed; 9 - an air electrode through which no energetic electrons passed.

RREA electrons leave behind about 29 thermal free electrons/cm of pathlength at 8 km altitude and about 18 electrons/cm at 12 km altitude (Rutjes et al., 2019). An avalanche of relativistic electrons (marked "4" in Figure 5.II) crosses the entire EE-volume in 1.5–3 μs. When runaway electrons traverse an E$_{th}$-volume (see Figure 5.II(6)), there is a high probability that discharge avalanches will develop and turn into streamers because most of the E$_{th}$-volumes are inside a E$_{str+}$-volume (see Figure 5.I(3)). The flash of streamers will form a front of current in approximately 30 ns, and the total duration of the streamer flash will be in the range of 150-300 ns. Thus, individual streamer flashes occur (see Figure 5.II(5)) and produce electromagnetic pulses with a radiation maximum in the VHF range of 30-40 MHz.

The avalanches of runaway electrons seeded with EAS act as an initiation wave moving at a speed close to the speed of light, which creates, at different points in the EE-volume, and in 1.5–3 µs, a "giant wave" of ordinary positive streamer flashes, which we described above (section 4.1). It can be said that a flash of a large number of runaway electrons "ignites" nearly simultaneously many of the $E_{th}$-volumes (see Figure 5.II(6)), thereby forming the NBE radiation front in the VHF. The motion of the phase wave of ignition of an exponentially growing number of streamer flashes is shown in more detail in Figure 7. (The numerical simulation is given in Kostinskiy et al., 2020).

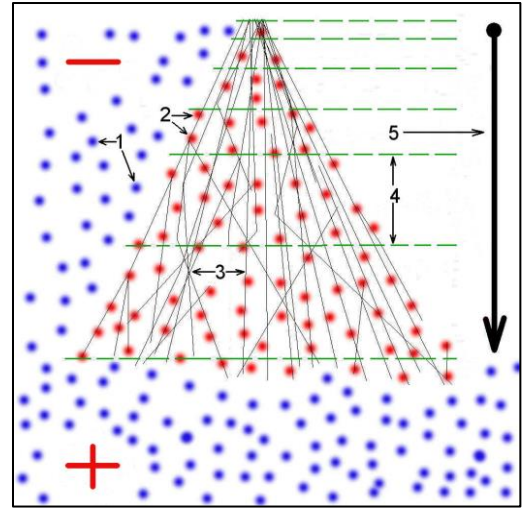

Figure 7. "Ignition" of the phase wave of streamer flashes by relativistic particles with a speed of $\sim 10^8$ m/s (fragment). 1 - an air electrode that has not been crossed by an energetic electron; 2 - an air electrode ($E_{th}$-volume) that has been crossed by an energetic electron; 3 - secondary EAS-RREA electrons; 4 - RREA runaway avalanche step; 5 - the phase wave of streamer flashes with a speed of $\approx 10^8$ m/s.

Thus, the NBE-IE Mechanism fulfills the conditions (c4, c5) of generation in several microseconds of a giant flash of positive streamers during the time of emission of NBEs without preliminary formation of a hot plasma channel of high conductivity. At the same time, the generation of this giant positive streamer flash is inextricably linked with avalanches of runaway relativistic electrons that were initiated by secondary electrons of EAS ($10^{15}$ eV < $\varepsilon_0$ < $10^{16}$ eV). Without EASs, avalanches of runaway electrons initiated by background cosmic rays cannot cross into most small air electrodes. On the other hand, with EAS only, even with an initial particle energy $\varepsilon_0 \approx 10^{17}$ eV, the EAS will not be able to initiate a sufficient number of streamer flashes to provide a large VHF signal that accompanies NBEs (Kostinskiy et al., 2020).

In constructing this part of the Mechanism, we used ordinary streamer flashes, which can move at reasonable speeds for streamers $v_{str} \approx 2\times10^5$-$3\times10^6$ m/s. Meanwhile the light speed of the relativistic runaway breakdown particles ignite the $E_{th}$-volumes along the paths of the relativistic particles, thereby giving the NBE an *apparent* speed close to the speed of light and satisfying conditions c6 and c7 (above, section 2).

### 5.1.4. Development of UPFs.

After the start of the streamer flashes, some streamers will be able to advance for distances of ~1-10 m because they are in more extensive $E_{str+}$-volumes, see Figure 5.III(12). During the movement of streamers, due to the ionization-heating instability, a network of strongly conducting plasma channels will appear in the plasma and remain after the streamers pass through (see Figure 5.III(7)). We will describe these strongly conducting plasma channels as "UPFs" since they develop like the UPFs reported in artificially charged aerosol clouds by Kostinskiy et al. (2015a; 2015b). It is important to note that the motion of one streamer flash will be able to simultaneously generate several UPFs, Figure 5.III(7). These UPFs will interact with each other after some time due to secondary positive streamers, as sketched in Figure 5.III(9) and as shown in experiments (Andreev et al., 2014) with a charged aerosol cloud, reproduced in Figure 8. The development time of the ionization-heating instability and the appearance of UPFs at atmospheric pressure is 0.2-0.5 µs; at altitudes where lightning is initiated, this time can increase by about a factor of ∼2-8 (Riousset et al., 2010). Thus, 5–20 microseconds (depending on height) after the onset of the VHF front of the NBE, the entire EE-volume will consist of small hot plasma formations or UPFs, ranging in length from a few centimeters to tens of centimeters. (See Supplemental Material S2 for more discussion of the ionization-heating instability and positive streamer-UPF transition.)

In order for the UPFs to "survive" for tens of microseconds, they must remain hot and conductive. Therefore, a current of no less than 0.2 A must flow through each hot channel UPF (Bazelyan & Raizer, 1998). UPFs that are close to each other will survive if they create an E ≥ 0.45-0.5 MV/(m·atm) between themselves (i.e., ≥ the minimum E needed to sustain positive streamer propagation) and if they exchange secondary positive streamers, similar to the way in which positive streamers support the development of space leaders in the crown of a negative long spark (Gorin & Shkilyov, 1976; Les Renardières Group, 1981). Due to the streamers between pairs of UPFs, the interaction is analogous to a small breakthrough phase (e.g., Figure 5.III(9,10)), which ends by combining and increasing the total length of the hot plasma channels in a process similar to a small "return stroke" (or step of a negative leader in a long spark) with an increase in current up to 5-15 A in a time of ∼1 µs. It is important to note that UPFs consist of a whole network of channels, where the current of small channels feeds large channels, helping them to survive longer: see Figure 9 (from Kostinskiy et al., 2015a), Andreev et al. (2014), and Figure 5.III(7). Thus, inside the EE-volume, many small channels are combined or merged into several large UPFs, Figure 5.IV(13). For long-term survival, each individual chain of UPFs must grow to such a length that the potential at its positive end reaches the 300-500 kV needed for starting a positive leader (Bazelyan & Raizer, 1998, 2000; Bazelyan et al., 2007a). (See Supplemental Material S3 for more discussion of the UPF-positive leader transition.)

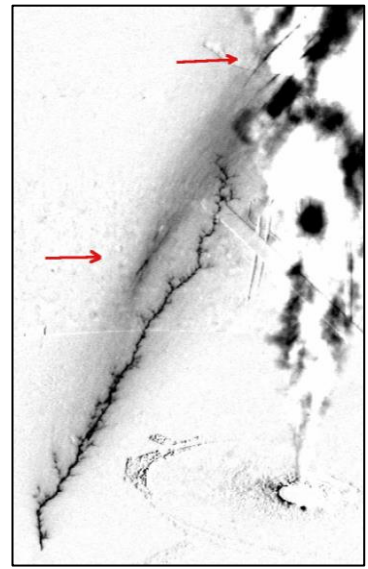

Figure 8. UPFs, formed after a positive streamer flash, interact with each other thanks to positive streamers. Negatively charged aerosol cloud. Infrared image with an 8 ms shutter speed. Two UPFs (indicated by red arrows) were born after the first streamer flash. UPFs interact through positive streamers. Upward positive leaders emerged and grew after the birth of UPFs (Adapted from *Andreev et al. 2014*).

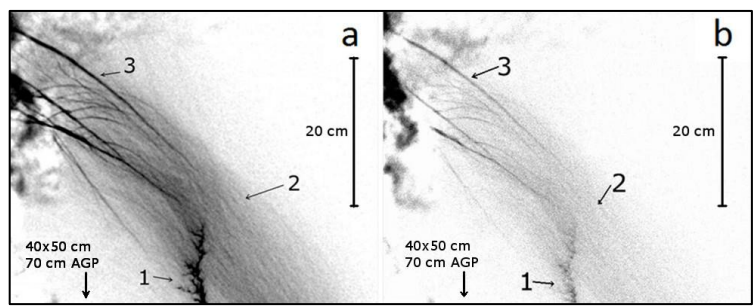

Figure 9.I. (Adapted from *Kostinskiy et al. 2015a*) Two consecutive infrared images (negatives) obtained with 6.7 ms exposure and separated by 2ms that show various discharge processes inside the cloud. Only flashes of scattered light, as opposed to distinct channels, were observed during this event in the visible range. Numbered features: 1: upper part of the upward positive leader (its lower part, developing in clear air, is outside the field of view of the IR camera), 2: streamer zone, 3: unusual plasma formation (UPF). AGP stands for "above the grounded plane."

### 5.1.5. Development of negative leaders.

According to the above sequence, many of the UPFs that have merged together in long chains and were able to become the "parents" of positive leaders have survived. As their positive leaders lengthen, the electric potential of their negative end will increase. A negative leader will be initiated from the negative end of a UPF chain when the potential at the negative end becomes approximately 1.5–2 times greater than the potential for initiating a positive leader (Gorin & Shkilyov, 1974, 1976; Les Renardières

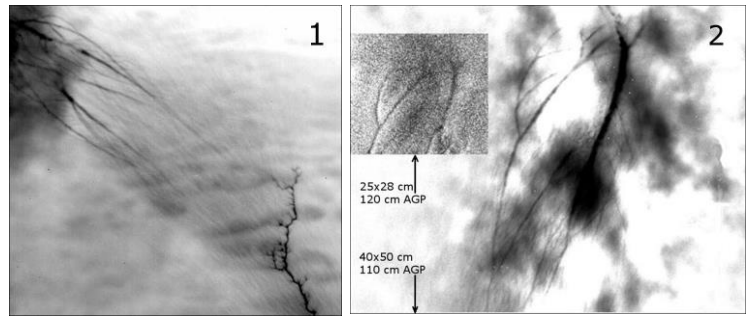

Figure 9.II. (1) Infrared image (negative) obtained with 6.7 ms exposure. The upper part of the upward positive leader (bottom right) and UPF (top left), both inside the cloud. (Adapted from *Kostinskiy et al. (2015a*). (2) (2) The visible image (top left) taken by a 4 Picos camera aimed at the center of the aerosol cloud at a height of 80-120 cm above the plane in which the grounded sphere equipped with current-measuring shunt. The camera was started by the current coming from the sphere through the shunt 400 ns after the initial streamer corona flash. Exposure time is 1 μs. The image is blurred since the UPFs are inside the cloud. Simultaneous IR image (right). Exposure time of the FLIR 7700M camera is 8 ms. It can be seen that the contours of the brightest UPF are similar in both images. (Adapted from *Andreev et al. 2014*).

Group, 1977, 1981). At this moment, the negative leader of Figure 5.IV(16) starts from the negative end of the network of the united UPFs, and the chain of UPFs turns into a "typical" bidirectional leader as shown in Figure 5.IV(13). The number of these bidirectional leaders in the EE-volume might vary in different flash initiations from a several tens to several hundred channels with a length of 10–50 meters. The number of bidirectional leaders will depend on the size of the initial EE-volume and the number of electron avalanches caused by runaway electrons passing through the EE-volume. (See Supplemental Material S4 for more discussion of the positive leader-bidirectional leader transition.)

Thus the NBE-IE Mechanism explains the IEC period of a lightning flash initiation as follows: First, the approximately simultaneous, 3-D production of many small UPFs is followed by the merging of small UPFs into UPF chains or networks; these events can occur during the IEC without large and strong discharges. Second, there is the development of positive and negative leaders from UPF networks, in preparation for the first IB pulse. The currents in the UPFs, UPF networks, and positive and negative leaders cause the change in E of the IEC detected at nearby fast antennas. We assume that the merging of two larger UPF chains or small bidirectional leaders that are formed from these UPF chains cause the IEC "enhancing events" described above and discussed in Marshall et al. (2014a; 2019). In these ways the Mechanism fulfills the IEC part of conditions c1 and c3.

### 5.1.6. Requirements for making IB pulses.

By definition the IEC starts with the IE and ends with the first classic IB pulse. As introduced above (idea i3) reported IEC durations average about 230 μs for CG flashes and 2700 μs for IC flashes. Modeling of the first or second classic IB pulse in three CG flashes found that the IBP peak currents were 30 - 100 kA, had durations of 30 - 50 μs, and had corona sheath charges (estimated line charges) of 0.2 – 1.2 C (Karunarathne et al., 2014). Thus, during the short time of the IE and IEC, the electron avalanches, positive streamers, UPFs, UPF networks, and bidirectional leaders must have liberated the 0.2 – 1.2 C of charge needed for the first classic IB pulse.

It is important to estimate the total length of UPF networks that will store the 0.2-1.2 C of charge, but the line charge density, $\omega$, of UPFs is unknown. For a long spark the measured $\omega = 0.05$-$0.07$ mC/m (Les Renardières Group, 1977, 1981), while for a well-developed negative lightning leader $\omega = 0.7$–$1.0$ mC/m (Rakov & Uman, 2003). To estimate the length of the UPF networks that combine to make an IBP, we use a linear charge density with a range of 0.07-0.7 mC/m and find lengths of 290-2900 m for 0.2 C and 1700-17000 m for 1.2 C. These lengths are much longer than the observed lengths (from high-speed video data) of 75-90 m for the first two IBPs in two negative CG flashes (Stolzenburg et al., 2013), so it seems impossible that a single bidirectional leader of 75-90 m could accumulate a charge of 0.2-1 C on its corona sheath. Therefore, the Mechanism postulates that the IEC is a three-dimensional process with parallel development of many UPF networks at once, and that each UPF network has its own three-dimensional structure (Kostinskiy et al., 2015a). In this way, during the IEC, the UPF networks can accumulate the required IBP charge. For an IBP length of roughly 100 m, the UPF networks would need 3-30 parallel elements for an IBP charge of 0.2 C and 17-170 parallel elements for an IBP charge of 1.2 C, depending on the UPF line charge density. Thus, the three-dimensionality of the UPF networks is fundamental for making an IBP. As the total length and volume of each UPF network increases and currents flow through them, the total charge in the corona sheath of each network increases; additional charge develops at the ends of the UPF network channels because the networks become polarized in the thundercloud electric field.

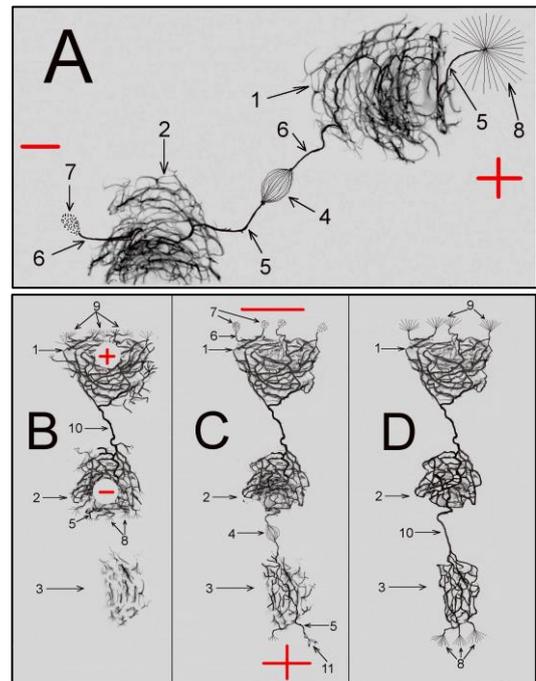

Figure 10 (sketch). A. Two plasma networks formed after the merger of UPFs interact with each other (the breakthrough phase of IBP#1). B. The "return stroke" phase of IBP#1 at which two plasma networks merge. C. The breakthrough phase of IBP#2. D. The "return stroke" phase of IBP#2. Numbered features: 1 - first plasma network, which was formed by the junction of many UPFs; 2 - second plasma network; 3 - third plasma network; 4 - the breakthrough phase contact of plasma networks; 5 - the negative leader; 6 - the positive leader; 7 - the streamer crown of the positive leader; 8 - flash of the streamer crown of the negative leader; 9 - flash of the streamer crown of the positive leader; 10 - the plasma channel of the "return stroke" phase; 11 - the streamer crown of the negative leader.

### 5.1.7. The first classic IB pulse.

According to the NBE-IE Mechanism, the first classic IBP occurs when two volumetric plasma systems, each consisting of a network of connected UPFs, send out bidirectional leaders that meet and connect, as schematically shown in Figure 10 A,B. The connection of the two networks involves a "breakthrough phase" (Figure 10 A) and a "return stroke" (Figure 10 B) (e.g. Jerauld et al., 2007) and accounts for the powerful flash of light associated with the first classic IB pulse (Stolzenburg et al., 2013, 2014; Campos & Saba, 2013). In this way the Mechanism fulfills conditions c1 and c8 by explaining the cause of the first classic IB pulse or IBP#1.

### 5.1.8. Subsequent classic IB pulses.

After IBP#1, E below the negative end of the two connected plasma networks (Figure 10 B.8) will be greatly enhanced by the "return stroke" of IBP#1 (and E above the positive end, Figure 10 B.9, will also be enhanced). As a result of the enhanced E below the negative end of IBP#1, one or more of the existing UPFs networks (Figure 10 B.3) in the enhanced E region will develop into a common network of hot plasma channels (Figure 10 C.3), at the ends of which a bidirectional leader will appear and connect to IBP#1 (Figure 10 C.4). This connection thereby makes the second classic IBP or IBP#2 with a new breakthrough phase (Figure 10 C.4) and "return stroke" (Figure 10 D.10). This sequence fits with the high-speed video and FA data of Stolzenburg et al. (2013; 2014), which showed that each new classic IBP extended the previous IBP channel. We note that unlike the generation of the first classic IBP, where the large E was caused by hydrodynamics and statistics, the generation of the second classic IBP is caused by the superposition of electric fields due to (1) E of the polarized plasma of IBP#1 and (2) E caused by hydrodynamics and statistics. The third classic IBP may occur in a similar way as IBP#2 providing the superposition of the electric fields remains sufficient for UPF networks to develop into a common network of hot plasma channels, at the ends of which a bidirectional leader will appear.

As mentioned in the Introduction (term t4), we defined the IB stage of the flash initiation as starting with the first classic IBP, and with an implied end time after the last IBP or at the transition to a negative stepped leader. We also mentioned that the processes causing classic IBPs and small IBPs were unknown and might be different. In the Mechanism the physical process of classic IBPs is the connection of two bidirectional leaders, each of which developed from a large three-dimensional network of UPFs (or from a three-dimensional network of hot plasma channels developed from the UPF network). In the Mechanism a different physical process causes both (i) small IBPs and VHF pulses between classic IBPs and (ii) the FA pulses and VHF pulses that occur during the IEC (before the first classic IBP): namely, the connection/merging of two UPFs, of a UPF to chain or network of UPFs, or a two chains (or small networks) of UPFs. We will group these three ways together under the name of "**preparatory mergers**." The largest of the preparatory mergers that occur during the IEC are the IEC enhancing events discussed above.

### 5.1.9. Transition to typical negative stepped leader.

As described above (section 5.1.5), negative leaders will be initiated from the negative ends of UPF networks (or a three-dimensional network of hot plasma channels developed from the UPF networks) when the potential at the negative ends becomes 1.5–2 times greater than the potential for initiating a positive leader. These positive leaders are shown in Figure 10(6). If there are other UPF networks (or another three-dimensional network of hot plasma channels developed from the UPF networks) nearby, then more IB pulses are possible; if not, then the negative stepped leader will survive if the UPF channel has become hot and conductive and the electric field is sufficient. When the negative leader (similar to the one shown in Figure 10D(8), in the process of developing bidirectional leaders, becomes self-propagating (i.e., becomes a negative stepped leader), then the process of forming the lightning channel is complete and the lightning initiation phase ends.

### 5.1.10.  Comparison of NBE-IE Mechanism to Data

Rison et al. (2016) studied in detail three positive NBEs that initiated IC flashes; the NBEs moved downward, and the following IB pulses moved upward. For each flash, immediately after the downward-moving positive NBE, the digital interferometer detected a series of scattered, low power, upward-moving events that Rison et al. (2016) identified as corresponding to the beginning of the IEC. These low power sources are shown in Figures 2a, 2b, and 3a in Rison et al. (2016).

In addition, their Figure 3c shows that these low power sources continued until the first IB pulse of the flash. We speculate that the low power sources were mergers of UPFs into UPF chains and UPF networks, with two UPF networks finally connecting to make the first IB pulse, as described above in sections 5.1.4, 5.1.5, and 5.1.7 and shown diagrammatically in Figures 10A and 10B.

Figure 3c of Rison et al. (2016) also shows that low power, upward-moving sources were almost continuously detected by the digital interferometer between the first and second IB pulses, and again between the second and third IB pulses. We speculate that the low power sources between each pair of IB pulses were also mergers of UPFs into UPF chains and eventually into a UPF network that caused the next IB pulse, when the newly-merged UPF network connected to the growing series of UPF networks (from previous IB pulses), as briefly described above in section 5.1.8. Figures 10B and 10C show diagrammatically the process leading to the second IB pulse of a flash via the merging of UPFs into the new UPF network (marked "3" in the Figures 10B and 10C). Figures 10C and 10D show diagrammatically that the second IB pulse occurs when the new UPF network ("3") connects to the pair of UPF networks that connected and made the first IB pulse.

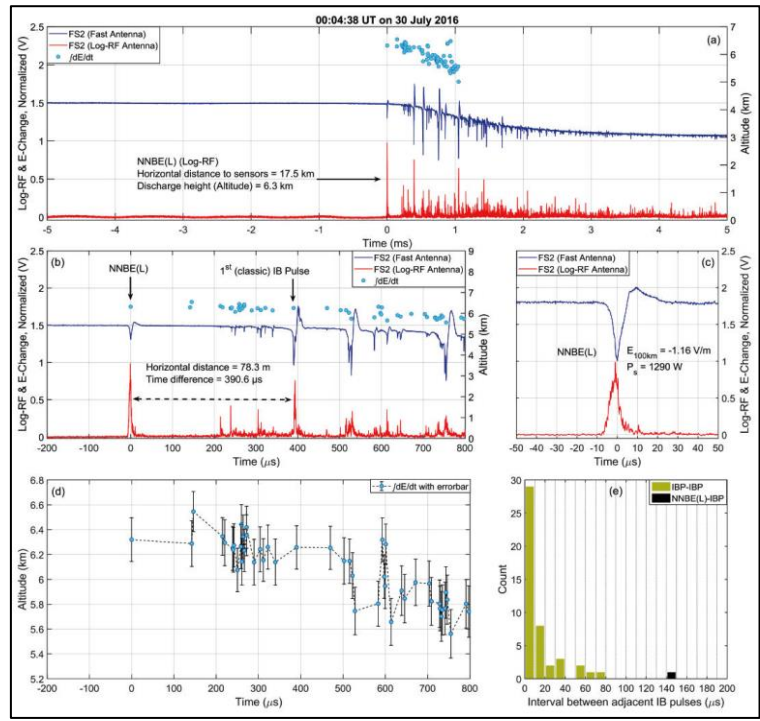

Figure 11. Example of a stronger-power NNBE (1290 W) that apparently initiated a -CG flash (called NNBE(L) in Bandara et al. (2019)). FA data (blue, uncalibrated linear scale) and Log-RF data (red, uncalibrated logarithmic scale) plotted as normalized voltage versus time (i.e., for each curve the largest peak-to-peak pulse amplitude is scaled to 1.0 V). $E_{100km}$ is the FA zero-to-peak amplitude (in V/m) of the NNBE(L), range-normalized to 100 km, while $P_S$ is the VHF power (in watts) of the NNBE(L). (a) Overview showing 10 ms of FA data and Log-RF data. Light blue dots represent altitudes (right-hand vertical scale) of FA pulses determined using $\int \frac{dE}{dt}$. Altitude of the NNBE(L) was 6.3 km. (b) Expanded view (1 ms) of first events in (a). (c) Expanded view (100 µs) of the NNBE(L). (d) FA pulse altitudes with error bars for the same 1 ms shown in (b). (e) Histogram of time intervals between adjacent FA pulses for the same 1 ms shown in (b) and (d). The time interval between the NNBE(L) and the next pulse location is shown in black in the histogram (*Bandara et al. 2019*).

Bandara et al. (2019) investigated negative NBEs (NNBEs) that initiated negative CG (-CG) flashes. NBE polarity is based on the polarity of the initial peak of the fast antenna NBE waveform using the physics convention of electric field polarity. We assume that positive streamer flashes are the underlying source of the NBE waveform. Hence, for -CG flashes, NNBEs move positive charge upward, and the following IBPs move downward and transition into a downward moving negative stepped leader.

In our opinion, the NBE-IE Mechanism for initiating lightning is qualitatively confirmed by the measurements of Bandara et al. (2019), as seen in Figure 11 and Figure 12. Figure 11 a shows a 10 ms overview the initial events in the -CG flash as detected by a fast antenna (FA) and a VHF power sensor (called Log-RF, bandwidth 186-192 MHz); note that the initial event in the flash, the NNBE, had the largest Log-RF power and that the Log-RF powers of the classic IB pulses were also relatively large. The blue dots in Figure 11 represent (z, t) locations of FA pulses; the (x, y, z, t) locations were determined using a time-of-arrival technique with an array of dE/dt sensors. Note that because of the amplitude scale used, some of the FA pulses with blue dot locations are not discernable in Figure 11. Figure 11c shows a 100 μs view of the NNBE: the FA bipolar pulse had an amplitude of -1.16 V/m (range-normalized to 100 km) and a duration of about 20 μs, while the VHF pulse (Log-RF) had a power of 1290 W and a duration of 15-17 μs. Note that the chaotic nature of the Log-RF pulse is consistent with the Mechanism's hypothesis that NBEs are an incoherent superposition of many positive streamer flashes.

Figure 11b shows the FA data, Log-RF data, and FA pulse altitudes for the first 800 μs of the flash; Figure 11d shows the FA pulse altitudes with z-error bars for the same 800 μs. The time between the NNBE and the first classic IB pulse was 390 μs (Figure 11b). From the point of view of the NBE-IE Mechanism, we assume in Figure 11b that the FA pulses (with blue dot locations) were caused by linking together UPFs into relatively long chains (long enough to make an FA pulse). The connections into UPF chains began only 140 μs after the IE, as indicated by the locations of the very weak FA and VHF pulses. Many chain mergers occurred in the time range from 220 to 340 μs after the IE. We assume that some of the merged UPF networks produced bidirectional leaders. When two of these leaders met and connected, the substantial "return stroke" current produced the first classic IB pulse. Note that the "preparatory mergers" from 220 to 340 μs occurred at short time intervals in the range of 2–15 μs (Figure 11d, e). Similar preparatory mergers occurred before the second classic IBP

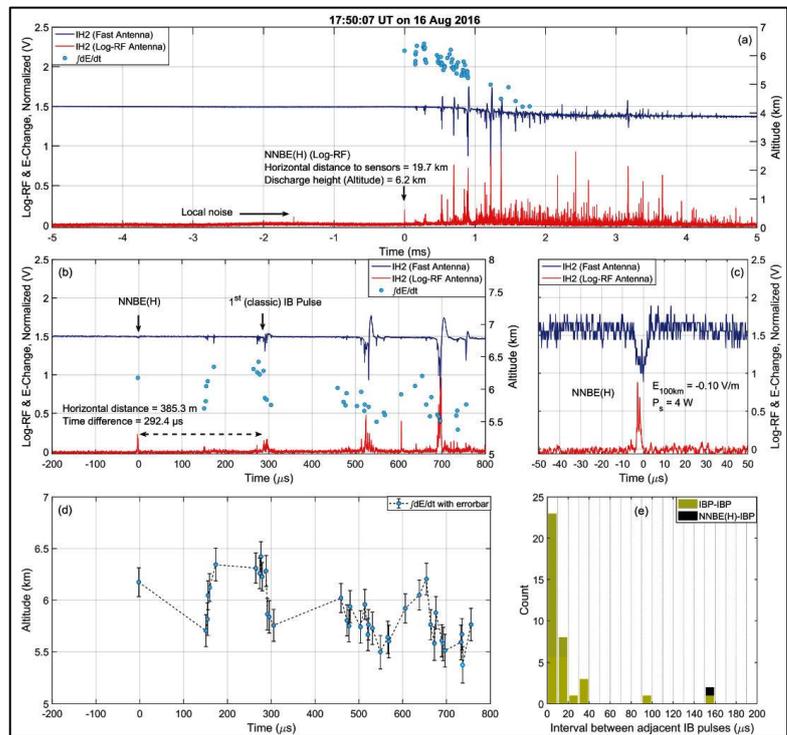

Figure 12. (*Bandara et al. 2019*). Similar to Fig.13, showing an example of a weaker-power NNBE (4 W) that apparently initiated a -CG flash (called NNBE(H) in Bandara et al. (2019)).

(IBP#2 at about 530 μs after the IE) and before the third classic IBP (IBP#3 at about 750 μs after the IE). The preparatory mergers of UPFs chains (or networks) before IBP#2 and IBP#3 might have involved only UPFs caused by the original avalanches/positive streamer flashes but might also have included a new UPFs caused by later avalanches/positive streamer flashes. Overall, the NBE-IE

Mechanism qualitatively fits with the data in Figure 11 of a strong NNBE initiation of a -CG flash reported by Bandara et al. (2019).

Figure 12 (Bandara et al., 2019) shows the initiation of a -CG flash by a weaker NNBE, and this initiation is also consistent with the NBE-IE Mechanism. The FA bipolar IE pulse had an amplitude of only -0.01 V/m (range-normalized to 100 km) and a duration of about 10 μs; the Log-RF pulse had a VHF power of 4 W and a duration of about 5 μs. Note that the FA pulse was not clearly bipolar; instead it was "more monopolar in nature," as in one NNBE reported by Rison et al. (2016). The NNBE was the first event in the flash and had a fairly large power, larger than the power of the first classic IBP, but much smaller than later classic IBPs. The first classic IBP (IPB#1) occurred about 290 μs after the NNBE; apparent preparatory mergers began 150 μs after the NNBE with more mergers in the 20 μs before IPB#1. Classic IBP#2 occurred 530 μs after the IE, and preparatory mergers began 80 μs before IBP#2. Classic IBP#3 occurred 700 μs after the IE; preparatory mergers occurred throughout the time between IBP#2 and IBP#3 and increased just before IBP#3. Immediately before each of the first three classic IBPs, the mergers occurred at short time intervals in the range of 2-10 μs (Figure 12d, e).

The short measured times (e.g., 5-150 μs, Figures 11 and 12) between the mergers might be accounted for by the close distances between the interacting channels. We can support this idea using a simple estimation. Consider that the average leader speed in the initial stage of lightning development is in the range $v_L \approx 0.02 - 0.1 \frac{m}{\mu s}$ (Rakov & Uman, 2003; Gorin & Shkilyov, 1976; Les Renardières Group, 1977, 1981). Further, consider that the average speed of leaders in the breakthrough phase is in the range $v_{L_{br}} \approx 0.1 - 0.3 \frac{m}{\mu s}$ (Rakov & Uman, 2003; Gorin & Shkilyov, 1976; Les Renardières Group, 1977, 1981). Hence, the distance between the plasma channels would be in the range $D_{ch} \approx 0.1 \frac{m}{\mu s} \cdot (5 - 150)\mu s \approx (0.5 - 15) \, m$. This general estimate is consistent with our previous estimates (section 5.1.2) of the distance between the air electrodes (4.6 m) from which the entire process of network formation begins.

The -CG flash in Figure 12 was initiated with a much weaker NNBE than the flash in Figure 11, but after the IE, the development of these two flashes seems quite similar. Both initiations seem to fit reasonably well with the NBE-IE Mechanism.

### 5.1.11. NBE-IE Mechanism for Precursor events and Isolated NBEs

Rison et al. (2016) described a kind of "short duration discharge," isolated in time and space, that they called "precursors (PCs)" since they "sometimes occur seconds before an IC discharge initiates at the same location." The two PCs shown in Rison et al (2016) had durations of 250 μs and 3 ms. The respective PCs had IEs with durations of < 1 μs and 2 μs and VHF powers of 10.8 dBW and 21.6 dBW. The IE powers of the PCs were 20-30 dBW greater that the other PC events. Another type of short duration discharge is an NBE that is isolated in time and space; originally all NBEs were thought to be isolated discharges (e.g., Willett et al., 1989). From the point of view of the Mechanism, it is likely that precursors and isolated NBEs develop with the NBE-IE Mechanism, but their EE-volumes have only small $E_{str+}$-volumes (with E sufficient for the development of positive streamers), so that the development of large UPF networks and bidirectional leaders cannot occur, thereby preventing IBPs from occurring. Without IBPs, the precursors and isolated NBEs cannot develop into full lightning flashes.

## 5.2. The Weak-IE Mechanism for lightning initiation

In this section, we describe how the Mechanism accounts for lightning initiation by short duration (≤ 1 µs), low VHF power (< 1 W) initiating events, as described by Marshall et al. (2019) and Lyu et al. (2019). We call this part of the Mechanism the Weak-IE Mechanism. These IEs are weaker than all of the NBE initiating events described above and are clearly not NBEs. As mentioned in the Introduction, recent measurements indicate that 88% of 26 nearby IC flashes were initiated by Weak-IEs (Lyu et al., 2019) while 96% of 868-CG flashes were initiated by Weak-IEs (Bandara et al., 2019), so lightning flashes are 10-25 times more likely to begin with Weak-IEs rather than NBE-IEs. Thus we can expect that conditions needed for the Weak-IE Mechanism will be more likely to occur in a thundercloud than the conditions for the NBE-IE Mechanism.

### 5.2.1 The first condition of Weak-IE Initiations.

In agreement with the above-cited observations, *the VHF power of the IE should be < 1 W.* From the point of view of the Mechanism, the magnitude of the VHF power of the IE essentially depends on the number of $E_{th}$-volumes (or "air electrodes") in the EE-volume that start avalanches. If there are relatively few $E_{th}$-volumes starting avalanches, then the VHF signal of the IE will be small. Compared to the NBE-IE Mechanism, the Weak-IE Mechanism must have fewer $E_{th}$-volumes involved or fewer relativistic particles to start avalanches or a combination of these two conditions.

### 5.2.2 The second condition of Weak-IE Initiations.

As described in section 5.1.6., *the IEC must collect 0.2 − 1.2 C of charge for the first classic IB pulse.* The Weak-IE Mechanism (like the NBE-IE Mechanism) assumes that the first classic IBP is caused by the connection of two bidirectional leaders which developed from two UPF networks. To have sufficient charge for the first classic IBP, the total charge (total system capacity) distributed on the corona sheaths of the two merging bidirectional leaders (and the plasma networks that these leaders support) must be 0.2 - 1.2 C. This total charge is needed even for the shortest duration IECs (of order 100 µs). The amount of charge that will be moved during the IBP breakthrough phase and "return stroke" must be accumulated during the IEC.

### 5.2.3 The Weak-IE Mechanism.

In order for these two conditions to be fulfilled and the lightning to be initiated by the same Mechanism proposed for NBE-IEs, it is necessary that much of the EE-volume must include large $E_{str+}$-volumes, see Figure 13.I. (The reason for this requirement will be given in the next paragraph.) Also, many fewer $E_{th}$-volumes (air electrode) are needed (of order $10^2$-$10^4$ versus $10^7$). Since the minimum electric field magnitude in the $E_{str+}$-volumes (≥ 0.45-0.5 MV/(m·atm)) is only about 50% larger than the average electric field in the EE-volume (0.28-0.35 MV/(m·atm)), the electric field "landscape" of the Weak-IE Mechanism is more easily (and more frequently) realized from statistical fluctuations than the E landscape needed for the NBE-IE Mechanism with its very large number ($10^7$) of $E_{th}$-volumes. This landscape should also be crossed by a sufficient number of relativistic runaway electrons to start avalanches and positive streamer flashes (Figure 13.I(5)), but not more than needed in the NBE-IE Mechanism.

The key difference between the Weak-IE Mechanism and the NBE-IE Mechanism is based on the $E_{str+}$-volumes. Because most of the $E_{th}$-volumes are located in the $E_{str+}$-volumes, streamer flashes starting in an $E_{th}$-volume will have very long streamer trajectories (tens of meters) (Figure 13.II(6)) because the trajectories will continue for the extent of the $E_{str+}$-volume (Figure 13.I(3,4)). Due to the long trajectories, the ionization-heating instability will produce many UPFs in each streamer flash (Figure 13.III(7)). A few microseconds after the Weak-IE, the UPFs will be connected into long UPF chains by their own secondary positive streamer crowns (Figure 13.III(9)) because inside these chains the electric field is higher than the propagation threshold of positive streamers $E_{str+}≥0.45-0.5$ MV/(m·atm). Inside each long chain of UPFs, a current flows in the range of 5-20 A. The average speed of each UPF chain or plasma channel, which survives and moves due to the current of positive streamers connecting them, will be about 1-2 cm/µs (Les Renardières Group, 1977). When UPF channels are combined or merged into a few longer chains, then they can move 3-6 m towards each other in 150 µs. If UPFs are quasi-uniformly distributed, then the chaining of UPFs into one large hot channel can occur in a series of current pulses, with a time between pulses of 1-3 µs. Each UPF or small UPF chain that merges into the main local UPF chain will produce a significant current pulse in the combined, highly conductive channels Figure 13.IV(11). We hypothesize that these mergers cause VHF pulses of different amplitudes depending on the lengths of the UPF chains that are merging. Eventually, connecting one more UPF chain with a long UPF chain will produce a single plasma channel that is several meters long, long enough that strong negative and positive streamer flashes occur at the ends of the combined channel and a bidirectional leader is born, Figure 13.IV(14,15). These streamer flashes will be similar to positive and negative flashes of a long spark (Kostinskiy et al., 2018, 2015b) and will produce a strong VHF signal. The merging of long UPF chains and/or the creation of a bidirectional leader may produce IEC enhancing events, which have a fast antenna pulse (from the long current) coincident with the VHF pulse.

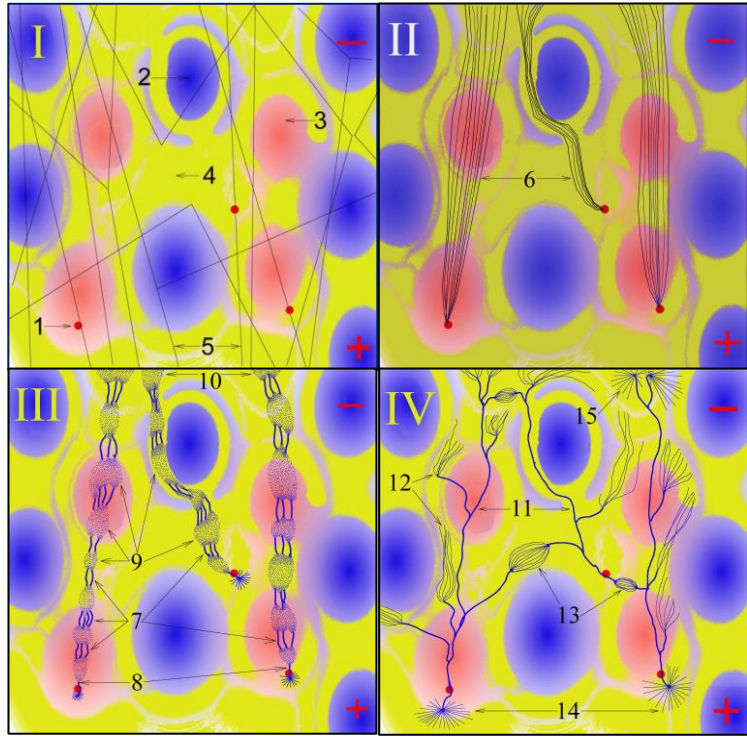

Figure 13. Possible initiation of lightning from the point of view of the Mechanism of "weak" NBEs and/or Initiating event (sketch). Numbered features: 1 - area with an electric field E≥ 3 MV/(m· atm); 2 - area with an electric field E< 0.45-0.5 MV/(m·atm); 3 - area with an electric field E ≥ 0.45-0.5 MV/(m·atm); 4 - area with an electric field E≈0.45-0.5 MV/(m·atm); 5 - runaway electron trajectories; 6 - long positive streamer flashes; 7,8 - UPFs; 9 - secondary streamer crowns connecting UPFs; 10 - positive crown ahead of UPF; 11 - hot highly conductive plasma channels; 12 - positive streamer crowns of positive leaders; 13 - positive streamer corona of two interacting large plasma channels; 14 - negative leader crown flash; 15 - positive leader crown flash.

Several three-dimensional plasma networks that create bidirectional leaders should develop close to each other. Then, as in the NBE-IE Mechanism, the first classic IBP, IBP#1, occurs when two of the bidirectional leaders connect to each other with a breakthrough phase and a "return stroke," Figure 13.IV(13). These events emit high power VHF pulses and large FA pulses of the first classic IBP.

The rest of the Weak-IE Mechanism is identical to the NBE-IE Mechanism. After the IBP#1, the E below the negative end of the two connected bidirectional leaders (Figure 13.IV.14) will be much enhanced by the "return stroke" of IBP#1, and one or more of the existing UPF networks in the enhanced E region will create bidirectional leaders and connect to IBP#1, thereby making the second classic IBP or IBP#2, etc. After enough IBPs have occurred to make a sufficiently conducting channel spanning a sufficient potential difference, a self-propagating negative stepped leader will begin, signaling the end of the lightning flash initiation.

### 5.2.4 Comparison of Weak-IE Mechanism to Data

Marshall et al. (2019) shows two examples of CG flashes with Weak-IEs, reproduced in Figure 14. Compared to the NBE-IE shown in Figures 11 and 12, the initiating event in Figure. 14a had a much smaller VHF power and a much shorter duration of 0.14 W and 1 μs; an expanded view (not shown) found no fast antenna (FA) pulse with the VHF initiating pulse. During the 130 μs IEC, there were many VHF pulses and only a few FA pulses; we assume that the VHF pulses were caused by UPFs connecting into UPF chains. There were two enhancing events (coincident FA and VHF pulses: -0.1 V/m & 0.18 W and -0.2 V/m & 0.14W) that occurred within 20 μs of the IE; these first events after the IE occurred much sooner than in the NBE-IE initiations discussed above. In the next 90 μs there were 7-10 small VHF pulses. In the next (last) 20 μs, there were many larger VHF pulses and one enhancing event (-0.2 V/m & 0.55W) leading up to the first classic IBP (IBP#1, -2.5 V/m & 3.0 W). These larger VHF pulses are consistent with the merging of longer UPF chains described in the Weak-IE Mechanism and similar to the late events in the IEC of NBE-IE flashes. The charge moment change of the IEC was 36 C m.

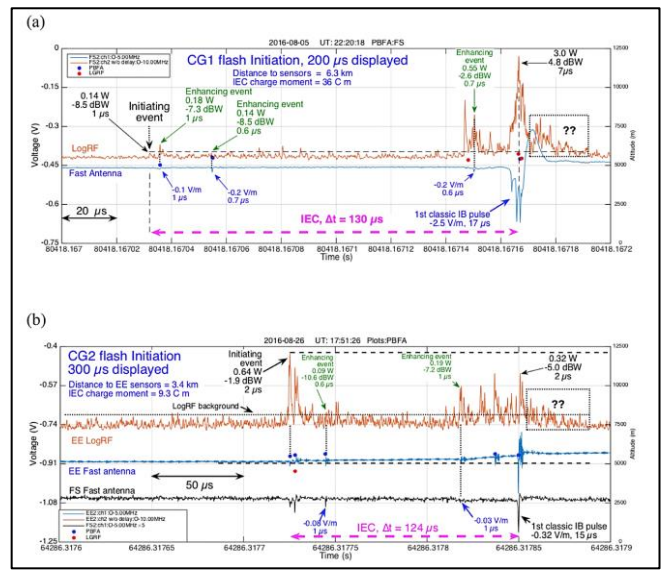

Figure 14. (a) First 200 μs of a –CG flash with Weak-IE (1 μs, 0.14 W). Fast antenna (FA) data in blue; VHF (LogRF) in red; distance from sensors to flash initiation was 6.3 km. The IEC lasted 130 μs. (b) First 300 μs of another –CG flash with Weak-IE (2 μs, 0.64 W); distance from EE sensors to flash initiation was 3.4 km. Similar to Figure 13a, but including data from an additional FA sensor (FS) shown in black (*Marshall et al. 2019*).

Figure 14b shows another Weak-IE CG flash initiation; it is similar to the initiation just discussed in many ways. However, the IE in Figure 14b was much stronger with VHF power and duration of 0.64 W and 2 μs and was coincident with a weak FA pulse. The IEC lasted 124 µs and had a charge moment change of only 9 C m. In the first 30 µs after the IE there were several substantial VHF pulses followed by one weak enhancing event (-0.08 V/m & 0.09 W). In the next 60 µs there were only a few VHF pulses. There were more than 20 relatively large VHF pulses in the last 40 µs before the first classic IBP (IBP#1, -0.32 V/m & 0.32 W).

Both flash initiations in Figure 14 seem to fit reasonably well with the Weak-IE Mechanism and, except for two differences, they are similar to the initiations via NBE-IE Mechanism shown in Figures 11 and 12. One of the differences is expected: the character of the IE itself (NBE-IE versus Weak-IE). The other difference is that the Weak-IE was immediately followed by VHF pulses while for the NBE-IE the VHF pulses did not begin for 140-150 µs. This difference can be explained by the Mechanism. The NBE-IE produces a large number of widely scattered (separated by 5-10 m), short UPFs. Apparently about 150 µs was needed before these UPFs had merged into short UPF chains, which could then merge and make observable VHF pulses. In the Weak-IE Mechanism, the long positive streamer flashes in the $E_{str+}$ volumes immediately developed multiple UPFs in each streamer flash. These UPFs quickly connected, and in 3-5 µs they developed lengths such that mergers would make detectable VHF pulses.

## 6. CONCLUSIONS

In this article, we have described a qualitative model of the physical processes of lightning initiation from the first event of the flash through the first few IBPs (Section 5). Our Mechanism suggests that lightning initiations develop as follows:

1. An initiating event (IE) begins the process of changing the non-conducting air into a conductor. The IE can be either weak or strong, called herein a Weak-IE or a NBE-IE, respectively. In both types of IEs, relativistic runaway particles seeded via Extensive Air Showers in a strong electric field, start classic electron avalanches in many small volumes in the thundercloud where the electric field, E > 3 MV/(m·atm). The three-dimensional (3D) group of electron avalanches causes multiple, nearly-simultaneous (synchronized), ordinary positive streamer flashes which radiate strongly in the VHF radio band. In this manner the Mechanism produces the IE and its characteristic VHF pulse.

2. The Initial E-Change (IEC) follows the IE in all (successful) flashes and involves Unusual Plasma Formations (UPFs). The UPFs develop within the positive streamer flashes via the ionization-heating instability process; the UPFs then merge together to form UPF chains or small networks. The electrical currents in the UPF networks make the relatively slow E-change of the IEC. Pairs of UPF chains also merge into longer and more complex chains. Then the chains form a three-dimensional network of hot plasma channels, which ultimately creates and supports the development of a bidirectional leader. The various mergers cause the weak VHF pulses and small fast antenna (FA) pulses seen during the IEC.

3. The first classic Initial Breakdown Pulse (IBP) ends the IEC and starts the IB stage. To produce the first classic IBP, two of the three-dimensional UPF networks (established during the IEC) must develop bidirectional leaders and then merge when their leaders contact each other. Each subsequent classic IBP is caused by the merger of a new three-dimensional UPF network to the string of previously connected UPF networks that caused the previous IBP(s). Each of these mergers has a common streamer zone, a breakthrough phase, and a miniature "return stroke" that causes a bright burst of light (in video data) coincident with the classic IBP (in FA data) and the high power pulse (in VHF data). After the series of classic IBPs, the lightning flash transitions to the well-known negative stepped leader phase.

As described above, the Mechanism is consistent with published data of lightning initiation for strong and weak NBE-IE flashes (Rison et al., 2016; Lyu et al., 2019; Bandara et al., 2019) and for Weak-IE flashes (Marshall et al., 2019) that do not begin with an NBE. The Mechanism can also be reasonably extended to explain small precursor-type events and isolated NBEs that are not IEs, in cases where the conditions for the IEC and/or the IBPs do not develop. Despite the qualitative character of the proposed Mechanism, it seems to us sufficiently concrete to test its main points in future experiments.

We conclude with a few important implications of the proposed Mechanism. First, the Mechanism assumes that the regions having sufficient E magnitudes to start ordinary positive streamer flashes are due to small-scale hydrodynamic instabilities and statistical variations of the electric field, and not due to the interaction of the electric field with hydrometeors. However, if hydrometeors are proven able to produce cloud volumes with fields $E_{th} \geq 3$ MV/(m·atm) necessary for starting streamers, then this circumstance will not significantly change the rest of the Mechanism. All the other components of the Mechanism would stay the same independent of the physical cause for the suitable large E magnitudes within small cloud volumes.

Secondly, in our Mechanism positive streamer flashes play the key role in making the IE, whether a Weak-IE or a strong NBE-IE. Although each streamer flash moves with a reasonable speed < 5 x $10^6$ m/s, the *apparent* motion during the IE can be much larger, 3-10x $10^7$ m/s. This fundamental difference between actual motion and apparent speed is because the streamer flashes are initiated along the paths of a group of relativistic charged particles which are moving essentially at the speed of light. In contrast, Rison et al. (2016) and Tilles et al. (2019) have postulated mechanisms for NBEs that initiate flashes based on Fast Positive Breakdown (FPB) moving downward and Fast Negative Breakdown (FNB) moving upward at speeds of 4-10x$10^7$ m/s. To us, such speeds for streamers do not seem reasonable, and, as far as we know, there is no experimental evidence for such high speeds of streamers in air at pressures of 0.3-1 atm.

The Mechanism proposes slightly different physical processes for the large-amplitude, long duration, "classic" IBPs (e.g., Weidman and Krider, 1979) and weaker, shorter IBPs (e.g., Nag et al., 2009). Weaker, shorter duration IBPs occur with the merging of two UPFs, a UPF with a UPF chain, or two UPF chains. Classic IBPs occur instead by the merging of two plasma networks, each of which evolved from a large UPF network. This result can help explain one mystery of the wide variation of activity observed in the IB stage.

Finally, an important feature of the proposed Mechanism is the fundamental three-dimensionality of the physical processes proposed, including the Initiating Event, rather than a single linear bidirectional leader. This approach makes it possible to explain short IEC times followed by powerful IBPs. Furthermore, because of the hypothesis of small-scale 3D variations in E, the Mechanism also readily explains the varied development of initiation events in different flashes, including the wide range of IEC durations and amplitudes (e.g., Marshall et al., 2014b), the wide range of IBP durations, inter-IBP times, IBP amplitudes, number of subpulses in the classic IBPs (e.g., Marshall et al., 2013; Stolzenburg et al., 2013; Stolzenburg et al., 2014; Bandara et al., 2019), and the seemingly random amplitude order of the classic IBPs (e.g. Smith et al., 2018). These variations among real lightning flashes are much harder to understand if initiation occurs a single smooth region of large E.

## Acknowledgments

We are grateful to Vladimir Rakov for a detailed discussion of the most controversial statements of the article, to Stanislav Davydenko, Alexander Litvak, Evgeni Mareev for discussions that showed the need to conduct more detailed numerical estimates of the synchronization in a significant volume of the cloud of many streamer flashes, to Nikolai Popov for a detailed discussion of the current state of the problem of ionization-heating instability, as well as to Gagik Hovsepyan and Ashot Chilingarian for the recommendation to use the EXPACS program and Andrei Vlasov and Mikhail Fridman for helping with the numerical calculation.

A. Yu. Kostinskiy participation in this research was supported by the Russian Science Foundation grant, 17-12-01439. T. C. Marshall and M. Stolzenburg participated in this research with support from a US National Science Foundation grant, AGS-1742930.

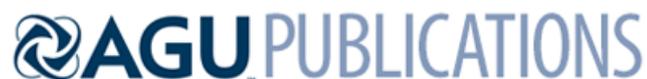

## The Mechanism of the Origin and Development of Lightning

## from Initiating Event to Initial Breakdown Pulses

**Alexander Yu. Kostinskiy[1], Thomas C. Marshall[2], Maribeth Stolzenburg[2]**

[1]National Research University Higher School of Economics, Moscow, Russia, kostinsky@gmail.com

[2]Department of Physics and Astronomy, University of Mississippi, University, Mississippi, USA,

marshall@phy.olemiss.edu , mstolzen@phy.olemiss.edu

**Contents of this file**

    Text S1 to S4

**Introduction**

The following sections S1 to S4 are supplementary explanatory text, providing more complete descriptions. References cited here are included in the References section.

## Text S1. Avalanche-streamer-transition.

As noted in pioneering works (Loeb, 1966; Phelps, 1974), positive streamers should play a key role in the origin and development of lightning, since the propagation of positive streamers in virgin air takes place with a smaller electric field $E_{str+} \approx$ 450-500 kV/(m·atm) than other plasma processes in ambient air, including the movement of negative streamers, which require a larger electric field, $E_{str-} \approx$ 1000-1200 kV/(m·atm) (Bazelyan & Raizer, 1998, 2000). Classical streamers are born due to avalanches of discharge electrons, which travel in an electric field $E_{th} \geq$ 3 MV/(m·atm) the distance necessary to fulfill the Meek's criterion (Raizer, 1991). Meek's criterion (depending on altitude/pressure and electric field strength, see Kostinskiy, Vlasov, & Fridman (2020)) implies that there must be areas in a thundercloud of at least 2-16 cm in size with a conventional breakdown electric field $E_{th} \geq$ 3 MV/(m·atm) so that $10^8$-$10^9$ electrons are concentrated in a volume $\leq$ 1 mm$^3$ (Raizer, 1991). With these parameters, the plasma is polarized in the external field $E_{str+} \geq$ 450-500 kV/(m·atm) and a condition of self-sustaining discharge is created in front of the head of the positive streamer (avalanche-streamer transition, Raizer, 1991).

Many studies suggest that cloud areas with a field $E_{th} \geq$ 3 MV/(m·atm), where an avalanche-streamer transition can occur, are formed by enhancement the field at the tips of one or more charged hydrometeors and/or due to hydrodynamic instabilities of the liquid phase hydrometeors (e.g., Loeb, 1966; Phelps, 1974). At present, such mechanisms for generating positive streamers cannot be excluded, but the small number of large hydrometeors in a thunderstorm cloud and their insufficient size drastically reduce the likelihood that such a mechanism for generating streamers is the main one (see section 4.3.1). A larger-scale hydrodynamic and statistical mechanism for enhancing the electric field seems more promising to us, but it currently requires further theoretical development and careful experimental verification (Trakhtengerts, 1989; Trakhtengerts et al., 1997; Mareev et al., 1999; Iudin et al., 2003; Trahtenhertz & Iudin, 2005; Iudin, 2017).

## Text S2. Ionization-heating instability and Streamer-UPFs transition.

The streamer plasma is cold and behind the streamer head the electrons disappear in 100–200 ns (Kossyi et al., 1992). Therefore, a fast mechanism for heating plasma streamers is necessary, since a plasma heated to temperatures above 2000-3000 K has a chance to "live" for several microseconds (Bazelyan & Raizer, 1998, 2000; Kossyi et al., 1992). The only currently known mechanism of transition of cold conducting plasma of streamers into a hot plasma of small channels is the well-developed mechanism of ionization-heating instability (Bazelyan & Raizer, 1998, 2000; Bazelyan et al., 2007a; Popov, 2009; da Silva & Pasko, 2012, 2013). In case of ionization-heating instability, the diameter of the channel through which the main part of the current flows decreases significantly (down to 50-100 μm, at a pressure of one atmosphere) compared to the original diameter of the streamer (about 1 mm), which allows the same current and the same external field to heat the air to high temperatures (Bazelyan & Raizer, 2000; Popov, 2009; da Silva & Pasko, 2012, 2013). To start the development of instability, it is necessary that either local air heating in the streamer channel increases the channel temperature by 10–20% (Raizer, 1991), or the electric field locally increases in the area of the streamer passage (Milikh et al., 2016). Candidates for such local heating are repeated passage of streamers along the same path (local heating) and/or statistical enhancement of the electric field due to uneven distribution of streamers in the volume of the streamer flash (Milikh et al., 2016). In the physics of a long spark, a similar process is called a streamer-leader transition (Les Renardières Group, 1977, 1981; Gorin & Shkilyov, 1974, 1976). However, in a long spark, a high voltage necessary for the development of the leader was originally applied to the place of origin of the leader on the electrode (stem) (Les Renardières Group, 1977, 1981; Gorin & Shkilyov, 1974, 1976). The presence of such a large local voltage is difficult to imagine in a virgin thundercloud. Therefore, relying on experiments (Kostinskiy et al., 2015a, 2015b), we will call this physical process a streamer-UPFs transition, meaning UPFs to be one or several hot plasma channels with a length of $\approx$ 5-30 cm formed after a streamer flash (or several flashes) in virgin air (with the polarization potential at the positive end of the UPF being lower than the potential for initiating a positive leader in a given external electric field).

Although UPFs and sparks are both hot, highly conductive plasma channels, there is another reason why we consider a UPF to be a separate plasma formation from a spark. With a classical spark discharge, the entire process of the formation of a short spark takes place in the breakdown fields $E_{th}$ $\geq$ 3 MV/(m·atm), regardless of whether the spark occurs via the Townsend avalanche development mechanism over distances of a few millimeters in length or via a streamer mechanism over longer intervals (Raizer, 1991). In contrast UPFs are produced in substantially subthreshold fields $E_{th}$ << 3 MV/(m·atm), where, almost always, the final mechanism of the transition and the gas heating necessary for the survival of the plasma is the ionization-overheating instability. The difference between the spark and a UPF is also visible in the threshold electric field. If the threshold for a short spark is well known($E_{th}$ $\geq$ 3 MV/(m·atm)) and understood as the field at which the ionization frequency begins to exceed the electron attachment frequency, then for ionization-overheating instability a reasonable range of electric fields is quite wide 0.5 $\leq$ Eth <3 MV/(m·atm), since the development of instability depends not only on the magnitude of the external electric field, but also on the scale of medium fluctuations. The cause may be an increase in heating, an increase in the electric field, or an increase in the concentration of electrons, since instability can develop from any link in the chain of events. For example: heating $\delta T\uparrow$ leads to a decrease in gas concentration $\delta N\downarrow$, a decrease in gas concentration leads to an increase in the ratio of electric field to molecular concentration $\delta(E/N)\uparrow$, an increase in E/N leads to a sharp increase in the electron energy and ionization frequency $\delta v_i\uparrow$, an increase in the ionization frequency leads to an increase in the number of electrons $\delta n_e\uparrow$, an increase in the number of electrons leads to an increase in the energy input to the gas $\delta(\sigma E^2)$, increased energy input to the gas again leads to heating $\delta T\uparrow$ and thus the process repeats and amplifies (Raizer, 1991; Bazelyan & Raizer, 1998, 2000; Bazelyan et al., 2007a, Popov, 2009; da Silva & Pasko, 2012, 2013). Note that this scenario of instability development occurs under a constant ambient subthreshold, electric field. The process can begin with any other parameter of the chain, for example, with a jump in the (subthreshold) electric field:

$$\delta E\uparrow \rightarrow \delta(E/N)\uparrow \rightarrow \delta v_i\uparrow \rightarrow \delta n_e\uparrow \rightarrow \delta(\sigma E^2)\uparrow \rightarrow \delta T\uparrow \rightarrow \delta N\downarrow \rightarrow \delta(E/N)\uparrow \ .$$

**Text S3. UPFs-positive leader transition.**

Hot, highly-conductive UPFs, which appear due to the streamer flash, cannot immediately be considered as leaders (especially bidirectional leaders). This is because in the average external electric field E<450-500 kV/(m·atm), the appearance of a positive leader from the plasma of the hot channel UPF requires that the plasma channel must reach a certain minimum (threshold) length, which, due to the polarization of the channel, will provide the potential necessary for the leader's development at the head of the positive end of the UPF (Bazelyan & Raizer, 2000; Bazelyan et al., 2007a). A single UPF, in order to "survive," must generate a sustainable positive leader. In order for the UPF to "live" until the leader starts from its positive end, a current must constantly flow through the UPF from the moment of birth, otherwise the electric field will be forced out of the plasma. The necessary scenario can be provided by chains of UPFs so closely spaced to each other that the electric field between them everywhere exceeds the propagation threshold of positive streamers $E_{str+}$ ≥450-500 kV/(m·atm). In this case, the ends of the channels of different UPFs will move towards each other at speeds of ≈ 2-6 cm/µs until several UPFs merge into a single channel. If the length of this longer single channel exceeds the threshold of initiation of a positive leader in a given electric field (Bazelyan et al., 2007a), then the positive leader will be self-sustaining and will lengthen in this field, thereby increasing polarization at the opposite negative end.

Another way of developing UPFs is realized if the average ambient electric field around UPFs exceeds the propagation threshold of positive streamers $E_{str+}$ ≥450-500 kV/(m·atm) and the length of such a field will be tens of meters. Interestingly, in this case, a streamer current supported by the ambient electric field will flow through *almost any* length of hot plasma (> 3-5 cm), and, consequently, the plasma channel will lengthen at a speed of 1-3 cm/µs, increasing potential at the negative end of the UPFs (Gorin & Shkilyov, 1976; Les Renardières Group, 1981). This option will be similar to the development of a short (≈ 5-15 cm), hot space-leader in the crown of the negative leader of a long spark (Stekolnikov & Shkilyov,1963; Gorin & Shkilyov, 1976; Les Renardières Group, 1981).

Since the process of development of UPFs is statistical and depends on the spatial size and intensity of the electric field, UPFs will not always grow to the initiation of a positive leader, and, therefore, the UPF plasma can disintegrate (decay). Finally, since the development of UPFs is another threshold process on the way from a streamer flash to a full-scale lightning flash, it seems to us that it should give a separate name, "UPF-positive leader transition," describing a positive leader "self-developing" from the positive end of a UPF when the external electric field is large enough.

**Text S4. Positive leader - bidirectional leader transition.**

When the potential at the negative end of the UPF exceeds 1.5-2 times the potential at the positive end of the UPF (Gorin & Shkilyov, 1974, 1976; Les Renardières Group, 1977, 1981), the electric field $E_{str-}$ ≥ 1000-1200 kV/(m·atm) will ensure the development of negative streamers at a distance of 20-100 cm before the UPF, then negative streamer flashes are initiated, and then, most likely, a small negative leader starts. After this, the bidirectional leader is likely to become self-sustaining and continue its development.

The development picture can become much more complicated if, as we suppose in the Mechanism, many close-lying UPFs, connected by positive streamers, are initiated at the same time. If the polarized three-dimensional UPF-plasma network has sufficiently large dimensions, then it will be able to form a slowly falling electric field in front of itself, which will increase the propagation length of positive streamers. To the best of our knowledge, such a plasma configuration in atmospheric pressure air in constant fields has not yet been considered theoretically or experimentally, so we cannot now predict in detail the evolution of such a network of UPFs. However, it can be assumed that as the UPFs merge and form highly conductive channels within the network, the highly conductive channels will concentrate the main current, and it is their polarization length that will determine the starting threshold of the positive leader from this UPFs network. After sufficient development of the positive leader, the necessary potential will build up at the negative end of the UPFs network to start a negative leader.

## References

Andreev, M. G., Bogatov, N. A., Kostinskiy, A. Yu., Makal'sky, L. M., Mareev, E. A., Sukharevsky, D. I., & Syssoev, V. S. (2014). First Detailed Observations of Discharges within the Artificial Charged Aerosol Cloud, XV International Conf. on Atmospheric Electricity, June 2014, Norman, OK, USA
https://www.nssl.noaa.gov/users/mansell/icae2014/preprints/Andreev_11.pdf.

Attanasio, A., Krehbiel, P. R., & da Silva, C. L. (2019). Griffiths and Phelps lightning initiation model, revisited, *Journal of Geophysical Research: Atmospheres, 124,* 8076-8094. https://doi.org/10.1029/2019JD030399

Babich, L. P., Bochkov, E. I., Kutsyk, I. M., Neubert, T. , & Chanrion, O. (2016). Positive streamer initiation from raindrops in thundercloud fields. *J. Geophys. Res. Atmos., 121*, 6393‐6403, https://doi.org/10.1002/2016JD024901

Bandara, S., Marshall, T., S., Karunarathne, N., Siedlecki, R., & Stolzenburg, M. (2019). Characterizing three types of negative narrow bipolar events in thunderstorms. *Atmospheric Research, 227,* 263-279, https://doi.org/10.1016/j.atmosres.2019.05.013

Bazelyan, E. M., & Raizer, Y. P. (1998). *Spark discharge*. Boca Raton, FL: CRC Press.

Bazelyan, E. M., & Raizer, Yu. P. (2000). *Lightning Physics and Lightning Protection*. IOP Publishing, Bristol.

Bazelyan, E.M., Raizer, Yu.P., & Aleksandrov, N.L. (2007a). The effect of reduced air density on streamer-to-leader transition and on properties of long positive leader. *J. Phys. D: Appl. Phys., 40,* 4133–4144, https://doi.org/10.1088/0022-3727/40/14/007

Bazelyan, E. M., Aleksandrov, N. L., Raizer, Y. P., & Konchakov, A. M. (2007b). The effect of air density on atmospheric electric fields required for lightning initiation from a long airborne object. *Atmospheric Research, 86*(2), 126‐138, https://doi.org/10.1016/j.atmosres.2007.04.001

Betz, H. D., Marshall, T. C., Stolzenburg, M., Schmidt, K., Oettinger, W. P., Defer, E., et al. (2008). Detection of in-cloud lightning with VLF/LF and VHF networks for studies of the initial discharge phase. *Geophys. Res. Lett., 35*(23), https://doi.org/10.1029/ 2008GL035820

Brothers, M.D., Bruning, E.S., & Mansell, E.R. (2018). Investigating the relative contributions of charge deposition and turbulence in organizing charge within a thunderstorm. *Journal of the Atmospheric Sciences, 75*(9), 3265-3284, https://doi.org/10.1175/JAS-D-18-0007.1

Campos, L. Z. S., & Saba, M. M. F. (2013). Visible channel development during the initial breakdown of a natural negative cloud-to-ground flash. *Geophys. Res. Lett., 40*, 4756–4761, https://doi.org/10.1002/grl.50904

Chapman, R., Marshall, T. C., Stolzenburg, M., Karunarathne, S., & Karunarathna, N. (2017). Initial electric field changes prior to initial breakdown in nearby lightning flashes. *Journal of Geophysical Research: Atmospheres, 122*, 3718-3732, https://doi.org/10.1002/2016JD025859

Colgate, S.A. (1967). Enhanced drop coalescence by electric fields in equilibrium with turbulence. *J. Geophys. Res., 72*, 479–487, https://doi.org/10.1029/JZ072i002p00479

da Silva, C. L., & Pasko, V. P. (2012). Simulation of leader speeds at gigantic jet altitudes, *Geophys. Res. Lett., 39*, L13805, https://doi.org/10.1029/2012GL052251

da Silva, C. L., & Pasko, V. P. (2013). Dynamics of streamer-to-leader transition at reduced air densities and its implications for propagation of lightning leaders and gigantic jets. *J. Geophys. Res., 118*(24), 13,561-13,590, https://doi.org/10.1002/2013JD020618

Dawson, G.A., & D.G. Duff (1970). Initiation of cloud-to-ground lightning strokes. *J. Geophys. Res., 75*, 5858, https://doi.org/10.1029/JC075i030p05858

Dubinova, A., Rutjes, C., Ebert, U., Buitink, S., Scholten, O., & Trinh, G. T. N. (2015). Prediction of lightning inception by large ice particles and extensive air showers. *Physical Review Letters, 115*(1), 015002, https://doi.org/10.1103/PhysRevLett.115.015002

Dye, J. E., Bateman, M. G., Christian, H. J., Defer, E., Grainger, C. A., Hall, W. D., et al. (2007). Electric fields, cloud microphysics, and reflectivity in anvils of Florida thunderstorms. *J. Geophys. Res., 112*, D11215, https://doi.org/10.1029/2006JD007550.

Dwyer, J.R. (2003). A fundamental limit on electric fields in air. *Geophys. Res. Lett., 30*, 2055, https://doi.org/10.1029/2003GL017781

Dwyer, J.R. (2007). Relativistic breakdown in planetary atmospheres. *Physics of Plasmas, 14* (4), 042901, https://doi.org/10.1063/1.2709652

Dwyer, J.R., & Uman, M.A. (2014). The physics of lightning. *Physics Reports, 534*, 147-241, https://doi.org/10.1016/j.physrep.2013.09.004

Gorin, B. N., & Shkilyov, A. V. (1974). Development of the electric discharge in long gaps subjected to positive voltage impulses. *Electricity, 2*, 29-38. (in Russian)

Gorin, B. N., & Shkilyov, A. V. (1976). Development of the electric discharge in long gaps rod - plane subjected to negative voltage impulses. *Electricity, 6*, 31-39. (in Russian)

Griffiths, R. F., & Phelps, C. T. (1976). A model for lightning initiation arising from positive streamer corona development, *J. Geophys. Res., 81*, 3671-76, https://doi.org/10.1029/JC081i021p03671

Gurevich, A. V., Milikh, G. M., & Roussel-Dupre, R. A. (1992). Runaway electron mechanism of air breakdown and preconditioning during a thunderstorm. *Physics Letters A, 165*, 463–467, https://doi.org/10.1016/0375-9601(92)90348-P

Gurevich, A. V., Zybin, K. P., & Roussel-Dupre, R. A. (1999). Lightning initiating by simultaneous effect of runaway breakdown and cosmic ray showers. *Physics Letters A, 254*, 79-87, https://doi.org/10.1016/S0375-9601(99)00091-2

Gurevich, A. V., & Zybin, K. P. (2001). Runaway breakdown and electric discharges in thunderstorms. *Physics-Uspekhi, 44*, 1119, https://doi.org/10.1070/PU2001v044n11ABEH000939

Iudin D. I., Trakhtengertz, V. Y., & Hayakawa, M. (2003). Fractal dynamics of electric discharges in a thundercloud. *Phys. Rev. E, 68*, 016601, https://doi.org/10.1103/PhysRevE.68.016601

Iudin D. I. (2017). Lightning-discharge initiation as a noise-induced kinetic transition. *Radiophysics and Quantum Electronics, 60*(5), 374-394, https://doi.org/10.1007/s11141-017-9807-x

Jerauld, J., Uman, M. A., Rakov, V. A., Rambo, K. J., & Schnetzer G. H. (2007). Insights into the ground attachment process of natural lightning gained from an unusual triggered lightning stroke. *J. Geophys. Res., 112*, D13113, https://doi.org/10.1029/2006JD007682

Karunarathna, N., Marshall, T. C., Stolzenburg, M., & Karunarathne, S. (2015). Narrow bipolar pulse locations compared to thunderstorm radar echo structure. *Journal of Geophysical Research: Atmospheres, 120*, 11,690-11,706, https://doi.org/10.1002/2015JD023829

Karunarathne, N., Marshall, T. C., Karunarathne, S., & Stolzenburg, M. (2020), Studying sequences of initial breakdown pulses in cloud-to-ground lightning flashes. *Journal of Geophysical Research: Atmospheres, 125*, https://doi.org/10.1029/2019JD032104

Karunarathne, S., Marshall, T. C., Stolzenburg, M., & Karunarathna, N. (2014). Modeling initial breakdown pulses of CG lightning flashes. *Journal of Geophysical Research: Atmospheres, 119*, https://doi.org/ 10.1002/2014JD021553

Karunarathne, S., Marshall, T.C., Stolzenburg, M., & Karunarathna, N. (2015). Observations of positive narrow bipolar pulses. *J. Geophys. Res. Atmos., 120*, 7128‐7143, https://doi.org/10.1002/2015JD023150

Kolmasova, I., Marshall, T., Bandara, S., Karunarathne, S., Stolzenburg, M., Karunarathne, N., & Siedlecki, R. (2019). Initial breakdown pulses accompanied by VHF pulses during negative cloud-to-ground lightning flashes. *Geophysical Research Letters, 46*, https://doi.org/10.1029/2019GL082488

Kossyi, I. A., Kostinsky, A. Y., Matveev, A. A., & Silakov, V. P. (1992). Kinetic scheme of the non-equilibrium discharge in nitrogen–oxygen mixtures. *Plasma Sources Sci. Technol., 1*(3), 207-220, https://doi.org/10.1088/0963-0252/1/3/011

Kostinskiy, A. Y., Syssoev, V. S., Bogatov, N. A., Mareev, E. A., Andreev, M. G., Makalsky, et al. (2015a). Observation of a new class of electric discharges within artificial clouds of charged water droplets and its implication for lightning initiation within thunderclouds. *Geophys. Res. Lett., 42*, 8165–8171, https://doi.org/10.1002/2015GL065620

Kostinskiy, A. Y., Syssoev, V. S., Bogatov, N. A., Mareev, E. A., Andreev, M. G., Makalsky, et al. (2015b), Infrared images of bidirectional leaders produced by the cloud of charged water droplets, *Journal of Geophysical Research: Atmospheres, 120*, 10,728–10,735, https://doi.org/10.1002/2015JD023827

Kostinskiy, A. Y., Syssoev, V. S., Bogatov, N. A., Mareev, E. A., Andreev, M. G., Bulatov, M. U., et al. (2018). Abrupt elongation (stepping) of negative and positive leaders culminating in an intense corona streamer burst: Observations in long sparks and implications for lightning. *Journal of Geophysical Research: Atmospheres, 123*, https://doi.org/10.1029/2017JD027997

Kostinskiy, A. Y., Marshall, T.C., Stolzenburg, M. (2019). The Mechanism of the Origin and Development of Lightning from Initiating Event to Initial Breakdown Pulses. *arXiv:1906.01033 v1 [physics.ao-ph]*.

Kostinskiy, A. Yu., Vlasov, A., & Fridman, M. (2020). Calculation of the dynamics of the initiation of streamer flashes that provide the NBE VHF signal profile and the VHF phase wave propagation velocity. *arXiv: 2005.12417 v1 [physics.ao-ph]*.

Le Vine, D. M. (1980). Sources of strongest RF radiation from lightning. *J. Geophys. Res., 85*, 4091– 4095, https://doi.org/10.1029/JC085iC07p04091

Les Renardières Group (1977). Positive discharges in long air gaps at Les Renardières. *Electra, 53*, 31–153.

Les Renardières Group (1981). Negative discharges in long air gaps at Les Renardières. *Electra, 74*, 67–216.

Loeb, L.B. (1966). The mechanisms of stepped and dart leaders in cloud-to-ground lightning strokes, *J. Geophys. Res.,* 71(20), 4711-4721, doi: 10.1029/JZ071i020p04711

Lyu, F., Cummer, S. A., Qin, Z., & Chen, M. (2019). Lightning initiation processes imaged with very high frequency broadband interferometry. *Journal of Geophysical Research: Atmospheres*, *124*, https://doi.org/10.1029/2018JD029817

Mareev E. A., Sorokin, A. E., & Trakhtengerts, V. Yu. (1999). Effects of collective charging in a multiflow aerosol plasma. *Plasma Physics Reports, 25*(3), 261-272.

Marshall, T. C., & Rust, W. D. (1991). Electric field soundings through thunderstorms. *J. Geophys. Res., 96*, 22,297‑22,306, https://doi.org/10.1029/91JD02486

Marshall, T.C., and M. Stolzenburg (1998), Estimates of cloud charge densities in thunderstorms, *J. Geophys. Res., D103,* 19,769-19,775.

Marshall, T. C., Stolzenburg, M., Maggio, C. R., Coleman, L. M., Krehbiel, P. R., Hamlin, T., et al. (2005). Observed electric fields associated with lightning initiation. *Geophys. Res. Lett., 32*, L03813, https://doi.org/10.1029/2004GL021802

Marshall, T.C., Stolzenburg, M., Karunarathne, S., Cummer, S., Lu, G., Betz, H.-D., & Briggs, M. (2013). Initial breakdown pulses in intracloud lightning flashes and their relation to terrestrial gamma ray flashes. *J. Geophys. Res.,* 118, 10907–10925, https://doi.org/10.1002/jgrd.50866

Marshall, T., Stolzenburg, M., Karunarathna, N. and Karunarathne, S. (2014a), Electromagnetic activity before initial breakdown pulses of lightning, *Journal of Geophysical Research*, 119, 12558–12574, https://doi.org/10.1002/2014JD022155

Marshall, T., Schulz, W., Karunarathna, N., Karunarathne, S., Stolzenburg, M., Vergeiner, C. & Warner, T. (2014b). On the percentage of lightning flashes that begin with initial breakdown pulses. *Journal of Geophysical Research: Atmospheres,* 119, 445–460, https://doi.org/10.1002/2013JD020854

Marshall T., Bandara, S., Karunarathne, N., Karunarathne, S., Kolmasova, I., Siedlecki, R., & Stolzenburg, M. (2019). A study of lightning flash initiation prior to the first initial breakdown pulse. *Atmospheric Research, 217*, 10-23, https://doi.org/10.1016/j.atmosres.2018.10.013

Milikh, G. M., Likhanskii, A. V., Shneider, M. N., Raina, A., George, A. (2016). 2-D model of the streamer zone of a leader. *J. Plasma Phys., 82*, 905820102, https://doi.org/10.1017/S0022377815001452

Nag, A., DeCarlo, B.A., & Rakov, V. A. (2009). Analysis of microsecond- and submicrosecond-scale electric field pulses produced by cloud and ground lightning discharges. *Atmospheric Research, 91*, 316–325, https://doi.org/10.1016/j.atmosres.2008.01.014

Nag, A., Rakov, V. A., Tsalikis, D., & Cramer, J. A. (2010). On phenomenology of compact intracloud lightning discharges. *Journal of Geophysical Research, 115,* D14115, https://doi.org/10.1029/2009JD012957

Nighan, W.L. (1977). Causes of thermal instability in externally sustained molecular discharges. *Phys. Rev. A, 15*(4), 1701-1720, https://doi.org/10.1103/PhysRevA.15.1701

Panchenko, V.Y., Zavalov, Yu.N., Galushkin, M.G., Grishaev, R.V., Golubev, V.S., Dubrov, V.D (2006). The development of turbulence in the active medium of a fast-flow gas-discharge laser, *Laser Physics, v.16*, no. 1, pp.40-51, https://doi.org/10.1134/S1054660X0601004X

Phelps, C .T. (1974). Positive streamer system intensification and its possible role in lightning initiation, *J. Atmos. Terr. Phys., 36*, 103-111, https://doi.org/10.1016/0021-9169(74)90070-1

Popov, N.A. (2009). Study of the formation and propagation of a leader channel in air. *Plasma Physics Reports, 35* (9), 785–793, https://doi.org/10.1134/S1063780X09090074

Raizer Yu., (1991). *Gas Discharge Physics*. Springer-Verlag, 449 pp.

Rakov, V. A., & Uman, M. A. (2003). *Lightning: Physics and Effects*. Cambridge Univ. Press, Cambridge.

Riousset, J. A., Pasko, V. P., & Bourdon, A. (2010). Air-density-dependent model for analysis of air heating associated with streamers, leaders, and transient luminous events. *J. Geophys. Res., 115*, A12321, https://doi.org/10.1029/2010JA015918

Rison, W., Krehbiel, P. R., Stock, M. G., Edens, H. E., Shao, X.-M., Thomas, R. J., et al. (2016). Observations of narrow bipolar events reveal how lightning is initiated in thunderstorms. *Nature Communications, 7*:10721, https://doi.org/10.1038/ncomms10721

Rutjes, C., Ebert, U., Buitink, S., Scholten, O., & Trinh, T. N. G. (2019). Generation of seed electrons by extensive air showers, and the lightning inception problem including narrow bipolar events. *Journal of Geophysical Research: Atmospheres, 124*, 7255–7269, https://doi.org/10.1029/2018JD029040

Smith, E. M., Marshall, T. C., Karunarathne, S., Siedlecki, R., & Stolzenburg, M. (2018). Initial breakdown pulse parameters in intracloud and cloud-to-ground lightning flashes, *Journal of Geophysical Research: Atmospheres, 123*, 2129–2140, https://doi.org/10.1002/2017JD027729

Stekolnikov I. S., & Shkilyov, A. V. (1963). The development of a long spark and lightning. Proc. of the Third Int. Conf. of Atmospheric and Space Electricity, Montreux, pp.466-481.

Stolzenburg, M., & Marshall, T. C. (2009). Charge structure and dynamics in thunderstorms, *Space Science Reviews, 137*(1–4) 355–372, https://doi.org/10.1007/s11214-008-9338-z

Stolzenburg, M., Marshall, T. C., Rust, W. D., Bruning, E., MacGorman, D. R., & Hamlin, T. (2007). Electric field values observed near lightning flash initiations. *Geophys. Res. Lett., 34*, L04804, https://doi.org/10.1029/2006GL028777

Stolzenburg, M., Marshall, T. C., Karunarathne, S., Karunarathna, N., Vickers, L. E., Warner, T. A., et al. (2013). Luminosity of initial breakdown in lightning, *Journal of Geophysical Research: Atmospheres, 118*, 2918–2937, https://doi.org/10.1002/jgrd.50276

Stolzenburg, M., Marshall, T. C., Karunarathne, S., Karunarathna, N., & Orville, R. E. (2014). Leader observations during the initial breakdown stage of a lightning flash. *Journal of Geophysical Research: Atmospheres, 119*, 12,198–12,221, https://doi.org/10.1002/2014JD021994

Stolzenburg, M., Marshall, T. C., & Karunarathne, S. (2020). On the transition from initial leader to stepped leader in negative cloud-to-ground lightning. *Journal of Geophysical Research: Atmospheres,* 125, https://doi.org/10.1029/2019JD031765

Tilles, J. N., Liu, N., Stanley, M.A., Krehbiel, P.R., Rison, W., Stock, M.G., Dwyer, J.R., Brown, R. & Wilson, J. (2019), Fast negative breakdown in thunderstorms, *Nature Communications*, 10:1648, https://doi.org/10.1038/s41467-019-09621-z

Trakhtengerts V. Yu. (1989). On the nature of electric cells in a thundercloud. *Reports of the USSR Academy of Sciences (Doklady Akademii Nauk SSSR), 308*(3), 584-586. (in Russian).

Trakhtengerts V. Yu., Mareev, E. A., & Sorokin, A. E. (1997). Electrodynamics of a convective cloud. *Radiophysics and Quantum Electronics, 40*(1-2), 77-86, https://doi.org/10.1007/BF02677826

Trakhtengerts, V. Y., & Iudin, D. I. (2005). Current problems of electrodynamics of a thunderstorm cloud. *Radiophysics and Quantum Electronics, 48*(9), 720–730, https://doi.org/10.1007/s11141-005-0116-4

Velikhov, E. P., Pis'mennyi, V. D., & Rakhimov, A. T. (1977). The non-self-sustaining gas discharge for exciting continuous wave gas lasers. *Soviet Physics Uspekhi-USSR, 20*(7), 586-602, https://doi.org/10.1070 / PU1977v020n07ABEH005445

Weidman, C. D., & Krider, E. P. (1979). The radiation field waveforms produced by intracloud lightning discharge processes. *Journal of Geophysical Research, 84*(C6), 3159‑3164. https://doi.org/10.1029/JC084iC06p03159

Willett, J.C., Bailey, J.C., & Krider, E.P. (1989). A class of unusual lightning electric field waveforms with very strong high-frequency radiation. *J. Geophys. Res. 94,* 16255. https://doi.org/10.1029/JD094iD13p16255.

Yuter, S. E., & Houze Jr., R. A. (1995). Three-dimensional kinematic and microphysical evolution of Florida cumulonimbus. Part I: Spatial distribution of updrafts, downdrafts, and precipitation. *Monthly Weather Review, 123*, 1921‑1940, https://doi.org/10.1175/1520-0493(1995)123<1921:TDKAME>2.0.CO;2